\documentclass[conference]{IEEEtran}
\pdfoutput=1
\IEEEoverridecommandlockouts

\usepackage{amsmath,amssymb,amsfonts}
\usepackage{algorithmic}
\usepackage{graphicx}
\usepackage{textcomp}
\usepackage{xcolor}

\usepackage{booktabs} 

\usepackage{caption}
\usepackage{subfig}
\usepackage{enumitem}

\DeclareCaptionLabelFormat{andtable}{#1~#2  \&  \tablename~\thetable}

\definecolor{brown}{HTML}{000000}
\definecolor{shadedgray}{gray}{0.93}

\begin{document}

\clearpage

\twocolumn

\pagenumbering{arabic}

\title{{LMKG}: Learned Models for Cardinality Estimation in Knowledge Graphs}

\author{\IEEEauthorblockN{Angjela Davitkova\textsuperscript{\textsection}}
\IEEEauthorblockA{
\textit{TU Kaiserslautern (TUK)}\\
Kaiserslautern, Germany \\
davitkova@cs.uni-kl.de}
\and
\IEEEauthorblockN{Damjan Gjurovski\textsuperscript{\textsection}}
\IEEEauthorblockA{
\textit{TU Kaiserslautern (TUK)}\\
Kaiserslautern, Germany \\
gjurovski@cs.uni-kl.de}
\and
\IEEEauthorblockN{Sebastian Michel}
\IEEEauthorblockA{
\textit{TU Kaiserslautern (TUK)}\\
Kaiserslautern, Germany \\
michel@cs.uni-kl.de}
}

\maketitle
\begingroup\renewcommand\thefootnote{\textsection}
\footnotetext{Equal contribution}
\endgroup

\begin{abstract}
\pdfoutput=1
Accurate cardinality estimates are a key ingredient to achieve optimal query plans. For RDF engines, specifically under common knowledge graph processing workloads,  the lack of schema, correlated predicates, and various types of queries involving multiple joins, render cardinality estimation a particularly challenging task. In this paper, we develop a framework, termed LMKG, that adopts deep learning approaches for effectively estimating the cardinality of queries over RDF graphs. We employ both supervised (i.e., deep neural networks) and unsupervised (i.e., autoregressive models) approaches that adapt to the subgraph patterns and produce more accurate cardinality estimates. To feed the underlying data to the models, we put forward a novel encoding that represents the queries as subgraph patterns. Through extensive experiments on both real-world and synthetic datasets, we evaluate our models and show that they overall outperform the state-of-the-art approaches in terms of accuracy and execution time. 

\end{abstract}

\pdfoutput=1

\section{Introduction}
\label{sec:introduction}
The capability of knowledge graphs (KG) to model structured data in a machine-readable way while having the possibility to be easily extended by interlinking data from various sources has contributed to their steadily increasing popularity.
Ranging from information retrieval applications to content-based recommendation systems, knowledge graphs are present in various domains, inspiring the perpetual initiation of new ideas and solutions. In recent years, techniques used for mining knowledge graphs have been widely researched and hugely impacted by deep learning models. Efforts for the improvement of RDF graph representation and embeddings by deep learning models have led to a promising performance in the widely studied tasks of question answering and link prediction~\cite{DBLP:journals/corr/abs-2002-00388}. Deep learning has also performed exceptionally well when considering the tasks of graph generation and processing~\cite{DBLP:conf/iclr/ShiXZZZT20,DBLP:conf/icml/YouYRHL18}. However, one area remains vaguely explored and that is the usage of deep learning models for query optimization in knowledge graphs. 

{\color{brown}
Intuitively, producing efficient query plans heavily relies on accurate cardinality estimates~\cite{DBLP:journals/pvldb/LeisGMBK015}. Although an RDF database can be seen as a single table composed of three columns (subject, predicate, object), traditional techniques used in relational databases have shown to perform poorly for SPARQL queries~\cite{DBLP:conf/edbt/Gubichev014,DBLP:conf/cikm/0003L11,DBLP:conf/icde/NeumannM11, DBLP:conf/www/StefanoniMK18}. 
The \textbf{challenges} in cardinality estimation come directly from the nature of RDF data and the lack of a rigid schema. \textit{First},  present correlation between the individual predicates renders the use of traditional cardinality estimation techniques, like histograms, inapt~\cite{DBLP:conf/icde/NeumannM11}. In other words, although the cardinality of two predicates independently may be quite selective, their co-occurrence can be quite common compared to other combinations---leading to an inaccurate estimate if independence is assumed. \textit{Moreover}, the SPARQL queries typically include many (self-) joins between RDF triples~\cite{DBLP:conf/edbt/Gubichev014}. Hence, to accurately estimate the cardinality, we need to go beyond the join uniformity assumption~\cite{DBLP:conf/cikm/0003L11, DBLP:conf/www/StefanoniMK18}. \textit{Finally},
it is not uncommon for SPARQL queries to contain more than one type of query patterns, like a query that exhibits both a star and a chain query pattern. Such queries
 that contain a compound of different patterns 
further contribute to the cardinality estimation challenges~\cite{DBLP:conf/cikm/0003L11, DBLP:conf/www/StefanoniMK18}.

} 

In this paper, we introduce \textit{LMKG}, \textit{a \underline{l}earned \underline{m}odel framework for cardinality estimation in \underline{k}nowledge \underline{g}raphs}. 
Given a knowledge graph, LMKG learns to estimate the cardinality of the most used types of queries (i.e., star and chain queries~\cite{DBLP:journals/pvldb/BonifatiMT17}).
Motivated by the recent research advancements for cardinality estimation in relational databases~\cite{LiuXYCZ15, Naru} {\color{brown}and the ability of neural networks to detect interconnections between variables}, with LMKG, we establish the problem of cardinality estimation of knowledge graph patterns as a deep learning problem. 
LMKG offers the creation of an \textit{unsupervised cardinality estimator} (LMKG-U) by employing autoregressive models with graph pattern encodings. 
By encoding the queries as graph patterns, LMKG also provides the possibility for creating a \textit{supervised cardinality estimator} (LMKG-S). LMKG efficiently learns the correlations between separate subgraph patterns and, as a result, provides distinctively accurate cardinality estimates. 
\textcolor{brown}{
To handle the challenge of high correlation between terms and the large number of (self-) joins, LMKG learns over relevant subgraph patterns and not independent terms or triples. To deal with the high number of patterns, LMKG provides an efficient sampling approach for generating relevant training data. 
} In addition to the most typically used encodings, a novel subgraph encoding coined \textit{SG-Encoding} is introduced. 
The SG-Encoding can incorporate various subgraph patterns while maintaining a compact representation. {\color{brown}Although SG-Encoding makes it possible to handle the challenge of composite patterns, the proof of concept and detailed evaluation is left for our future work.}

\subsection{Contributions and Outline}
The main contributions of this paper are:
\begin{enumerate}
	\item We formulate the problem of cardinality estimation in knowledge graphs through the lenses of supervised and unsupervised deep learned models.
	\item To tackle the problem of cardinality estimation in knowledge graphs, we develop a framework called LMKG that includes models of different types that can be tailored to a specific dataset or a sample workload. 
	\item To featurize  subgraph patterns and provide them as input in the models, we explore different encodings including our newly introduced           \textit{SG-Encoding}. 
	\item We report on a comprehensive experimental study, evaluating the LMKG framework against the state-of-the-art approaches, and finally, objectively discuss the challenges of learned knowledge graph cardinality estimation.
\end{enumerate}

The remainder of the paper is organized as follows. Section~\ref{sec:related_work} discusses related work. Section~\ref{sec:notation-problemStatement} provides the necessary notation and formulates cardinality estimation as a supervised and unsupervised problem. Section~\ref{sec:frameworkOverview} sketches the high-level outlook of LMKG and explains the comprising phases. The considered query types and the proposed encoding strategies are explained in Section~\ref{sec:join-queries}. Section~\ref{sec:lmkgUsedModels} details on the deep learning models used in LMKG. 
In Section~\ref{sec:appr_mode_overview_new} we elaborate on practical insights gained from the development of LMKG. Section~\ref{sec:experiments} reports on the results of the experimental evaluation, while, eventually, Section~\ref{sec:conclusion} concludes the paper.

\pdfoutput=1

\section{Related Work}
\label{sec:related_work}
\noindent
\textbf{Cardinality Estimation in Knowledge Graphs:}
Early work on cardinality estimation in knowledge graphs focuses on introducing new estimation algorithms that either use statistics that are gathered for the properties of the ontology~\cite{DBLP:journals/sigmod/ShironoshitaRK07} or statistics for the summary of the graph patterns~\cite{DBLP:conf/www/MadukoASS07}. The Jena ARQ optimizer~\cite{DBLP:conf/www/StockerSBKR08} uses pre-computed statistics and single attribute synopsis for estimating the join selectivity. However, the introduced estimation functions assume independence between the attributes which leads to underestimations. Similarly to the Jena ARQ optimizer, RDF-3X~\cite{DBLP:journals/vldb/NeumannW10} does not consider  correlation between  predicates. Vidal et al.~\cite{DBLP:conf/esws/VidalRLMSP10} suggest that basic graph patterns can be partitioned into groups that share one common variable, so called star-shaped groups. They use a sampling technique and introduce a cost model to estimate the cardinality of the star-shaped groups. 
Neumann and Moerkotte~\cite{DBLP:conf/icde/NeumannM11} introduce a synopsis called characteristic sets which they use as a basis for estimating the cardinality of SPARQL queries, again focusing on star-shaped queries. Gubichev et al.~\cite{DBLP:conf/edbt/Gubichev014} extend the statistics captured by the characteristics sets and use them for performing join ordering in SPARQL queries. 
Jachiet et al.~\cite{jachiet:hal-01524387} use statistics, mainly focused on the predicates. Huang and Liu~\cite{DBLP:conf/cikm/0003L11} propose combining two methods, Bayesian networks capturing the joint probability distribution over correlated properties for star query patterns and a chain histogram for chain query patterns. Presto~\cite{DBLP:journals/corr/abs-1801-06408} stores statistics of common subgraph patterns, relying on the presence of bound variables in the query. Stefanoni et al.~\cite{DBLP:conf/www/StefanoniMK18} represent the RDF graph in a more compact manner and use the created graph summaries for cardinality estimation. 
G-Care~\cite{gCare}, a recent benchmarking framework, compares existing approaches for cardinality estimation in graphs. They consider the state-of-the-art approaches in the area of knowledge graphs and additionaly, adapt existing cardinality estimators used in relational databases for graphs.

\noindent
\textbf{Learned Approaches for Cardinality Estimation in Relational DBMS:}
Recently, the usage of machine learning models for the improvement of traditional database components has expanded, ranging from applications in query optimization, approximate query processing or even complete system enhancement.
An early work, Leo~\cite{LEO} computes adjustments of the optimizer’s statistics and cardinality estimates later used during query optimizations, by monitoring previously executed queries.
Liu et al.~\cite{LiuXYCZ15} provide an effective neural network selectivity estimation for all relational operators, by taking a bounded range of each column as input. Similarly, MSCN~\cite{MSCN}, a multi-set convolutional network represents relational query plans with set semantics to capture query features and true cardinalities.
Dutt et al.~\cite{DuttWNKNC19} provide an application of neural networks and tree-based ensembles for selectivity estimation of multi-dimensional range predicates. 
For overcoming the issue of misestimates, 
Woltmann et al.~\cite{LocalDeepLearningModels} suggest a local-oriented approach. Hayek and Shmueli~\cite{LearningContainmentRates} propose the usage of estimated containment rates for improving query estimates. 
Exploration of simple deep learning models~\cite{SimpleModels} and pure data-driven models~\cite{DeepDB}, have also been proven efficient for cardinality estimation and beyond. Unlike the previous work, Naru~\cite{Naru} is an unsupervised data-driven synopsis that achieves high accuracy cardinality estimation in relational databases with adapted deep autoregressive models by introducing a new Monte Carlo integration technique called progressive sampling.
The recently proposed NeuroCard~\cite{NeuroCard} approach, independently developed from our work, extends the idea of autoregressive models for a join cardinality estimator over an entire database and has the potential to be applied on KGs. Deeper investigation on the applicability to cardinality estimation for KG queries is part of our future work.
Similarly, Hasan et al.~\cite{MultiAttributeCE}, suggest the usage of both autoregressive models and supervised deep learning models for accurate cardinality estimation.
Others~\cite{DQ, REJOIN2, Neo, StateRepresentationRL}, shift the focus from an explicit modeling of cardinalities to optimal plan generation by applying reinforcement learning.

\noindent
\textcolor{brown}{\textbf{Learning in Graphs:} 
Although not for cardinality estimation over knowledge graph queries, deep learning for KGs has been widely researched~\cite{DBLP:journals/corr/abs-2002-00388}.
Since we need to represent the subgraphs efficiently, the work of KG embedding is of particular importance for us. 
For instance, Hamilton et al.~\cite{DBLP:conf/nips/HamiltonYL17} propose an inductive approach that uses node features to learn an embedding function. TransE~\cite{DBLP:conf/nips/BordesUGWY13} and TransH~\cite{DBLP:conf/aaai/WangZFC14} create term embeddings by following the translation principle. However, generating the node embeddings in the presence of unbound terms has not been discussed. Additionally, this work mainly focuses on term or triple and not on a subgraph representation.
In a more related research, GraphAF~\cite{DBLP:conf/iclr/ShiXZZZT20} and MolecularRNN~\cite{DBLP:journals/corr/abs-1905-13372} focus on generating molecular graphs with the use of deep learning. 
Although efficient for generating molecules, this work does not allow the estimation of the individual term densities. To create our encoding, we build on their idea for subgraph representation.
}

\pdfoutput=1

\section{Notation \& Problem Statement}
\label{sec:notation-problemStatement}
An RDF\footnote{https://www.w3.org/RDF/} knowledge graph $KG$ is a finite set of triples. Each triple $t$ is constructed out of three terms $(s, p, o)$, corresponding to a subject, a predicate, and an object. Every term is uniquely identified with a URI where the objects can be literals (e.g., strings, integers). More specifically,  the subject is a resource or a node in the graph, via which a predicate forms a relationship to another node or a literal value, called an object. A standard query language for accessing RDF stores is SPARQL\footnote{https://www.w3.org/TR/sparql11-overview/}. SPARQL is based on matching graph patterns and there exist various types of patterns that a query can have (e.g., star, chain). A SPARQL query can have variables that are not bound to a specific term (e.g., $?x$).

Let us consider a knowledge graph $KG$ with triple patterns $t_i \in KG$ of the form $(s_i,p_i,o_i)$ with domains $s_i \in S, p_i \in P \text{ and } o_i \in O$. We consider  star and chain queries with an arbitrary number of joins and unbound variables. Let $qp$ be a SPARQL query consisting of a graph pattern $\{t_1,...,t_j,...,t_k\}$, where $k > 1 \land t_j \in KG$ and every triple $t_j$ can have an arbitrary number of unbound variables. The cardinality $card(qp)$ represents the number of graph patterns from the knowledge graph $KG$ that match the query graph pattern ${qp}$, i.e., the result size of $qp$. We aim at developing an estimator $est(qp)$ whose predicted value will be as close as possible to the real cardinality $card(qp)$. As an estimator, we explore the usage of \textit{supervised} and \textit{unsupervised} models.

\noindent
\textbf{Supervised estimator} As a supervised estimator, we investigate the usage of a deep neural network. The neural network will receive as input the query pattern $qp$ and will produce as output the estimated cardinality $est(qp)$. 

\noindent
\textbf{Unsupervised estimator} As an unsupervised estimator, we investigate the usage of an autoregressive model. 
These models decompose the joint density into $n$ conditional probabilities, where $n$ is the number of terms used to train the model. 

\section{Framework Overview}
\label{sec:frameworkOverview}

\begin{figure}[!t]

 \center
  \includegraphics[width=0.47\textwidth]{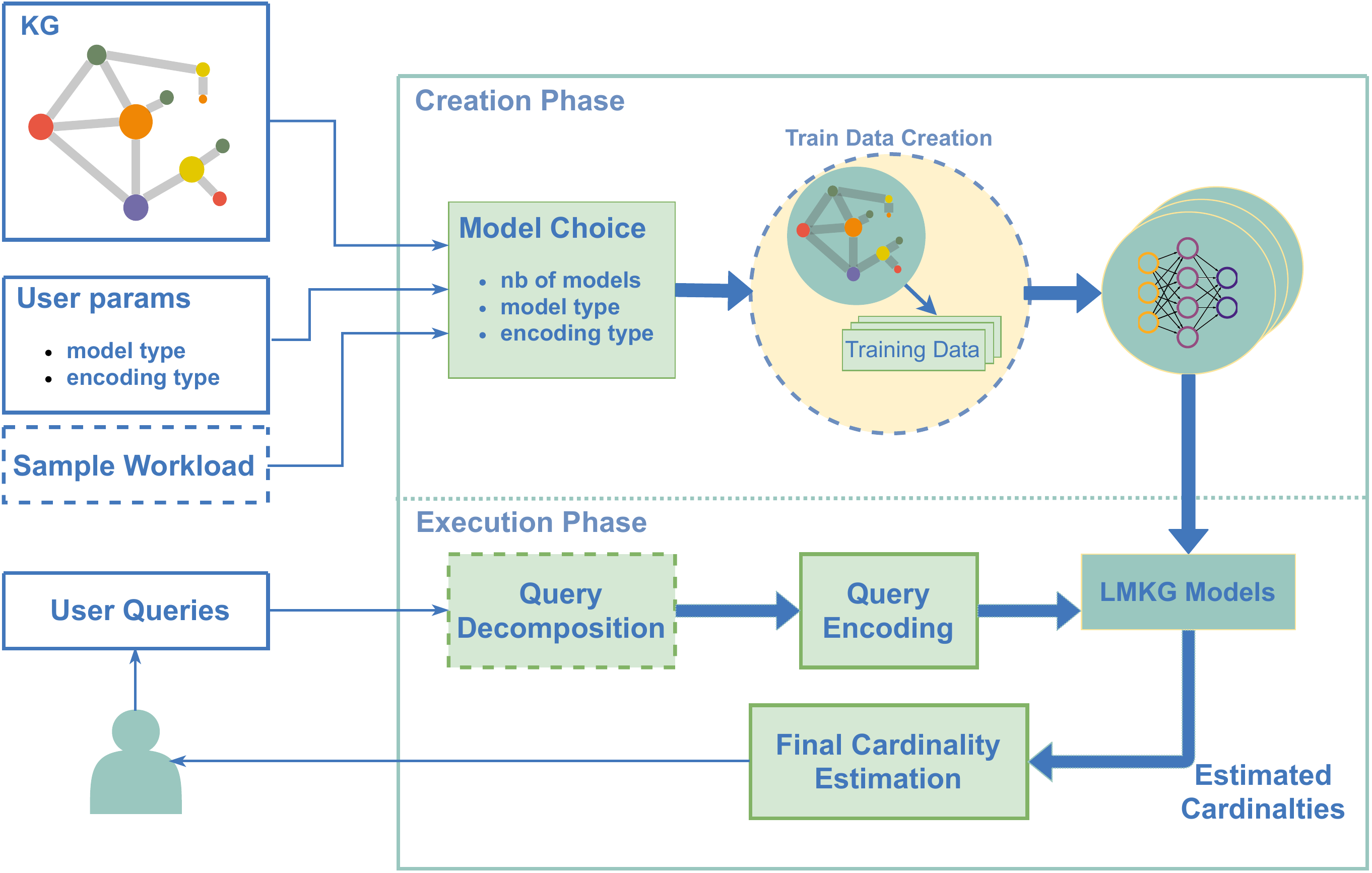}
  \caption{LMKG framework overview}
  \label{fig:figure-system-overview}

\end{figure}

As the name suggests, the LMKG framework represents a compound of several models. It comprises two phases that are depicted in \figurename~\ref{fig:figure-system-overview}. In the first phase, labeled the \textit{creation phase}, the initial step is to decide on the models that will be created, then generate the adequate training data if there is no sample workload available and, ultimately, train and tune the chosen learned models. The second phase, called the \textit{execution phase}, encompasses the user-system interaction, indicating the start of the querying process. Below, we delineate the tasks that are performed in the two phases.

\noindent
\textbf{Model choice}
\label{model-choice}
LMKG allows the creation of different models that can be tailored to a specific type of query.
Under the assumption that there is no existing query workload, initially, LMKG tries to create the most optimal models considering both the memory and the accuracy. However, if a sample query workload and a defined budget are provided, based on the workload statistics, LMKG can decide which models have a higher priority. 
Additionally, the number of models can  also be tuned by the user without a query workload. With our newly proposed encoding strategy (Subsection~\ref{encoding-strategy-our}), the user also has the possibility to create only one model capable of handling different types of queries (e.g., both star and chain queries). 
Having a single model that captures all query types and sizes may lead to larger errors. Therefore, we propose various model grouping strategies.
If a change in the workload of queries is detected during the execution phase, a new model may be created, or an existing model may be dropped.

\noindent
\textbf{Training data creation}
\label{trainign-data-creation}
Once the learned models have been decided upon and there is no sample workload available, the next step is to create the training data. 
LMKG creates the training data for the cardinality estimation models from the provided knowledge graphs, described in Section \ref{ssec:training_data_creation_appr_overview}.
When LMKG trains in a supervised manner, the training data consists of different graph patterns tailored for specific knowledge graph queries and their assigned cardinalities. It is important to note that the graph patterns can include unbound variables. For the unsupervised approach, the training data includes only the bound graph patterns since the model can estimate the conditional probabilities of the individual terms and use them for queries involving patterns with unbound variables. 

\noindent
\textbf{Training}
\label{creation_phase:training}
The training step uses the training data, either generated in the previous step or provided as a sample workload, to create one general model for multiple query types or several grouped models, each tailored for answering specific types and/or sizes of queries. 
The training step involves transforming the different graph patterns into features suitable for a deep learning model. This process depends on the chosen model. 

\noindent
\textbf{Querying}
\label{execution_phase:querying}
Once the model creation phase is finished, the models can be used from the user, indicating the start of the execution phase. Given a user-specified query, the task of LMKG is to provide an estimate of the query cardinality. As depicted in \figurename~\ref{fig:figure-system-overview}, if a query is of a specific type and size, which is already learned by one of the models, we can directly use the model to estimate its cardinality. However, if this is not the case and the query contains multiple patterns, it is forwarded to the Query Decomposition step, where it is decomposed into query patterns that are suitable for the existing models.
The query subgraphs are then forwarded to the encoding process where each pattern is encoded according to the model.
The featurizers then forward the inputs through the models and receive a prediction for the cardinality.

\begin{figure*}[!t]

 \center

  \includegraphics[width=0.9\textwidth]{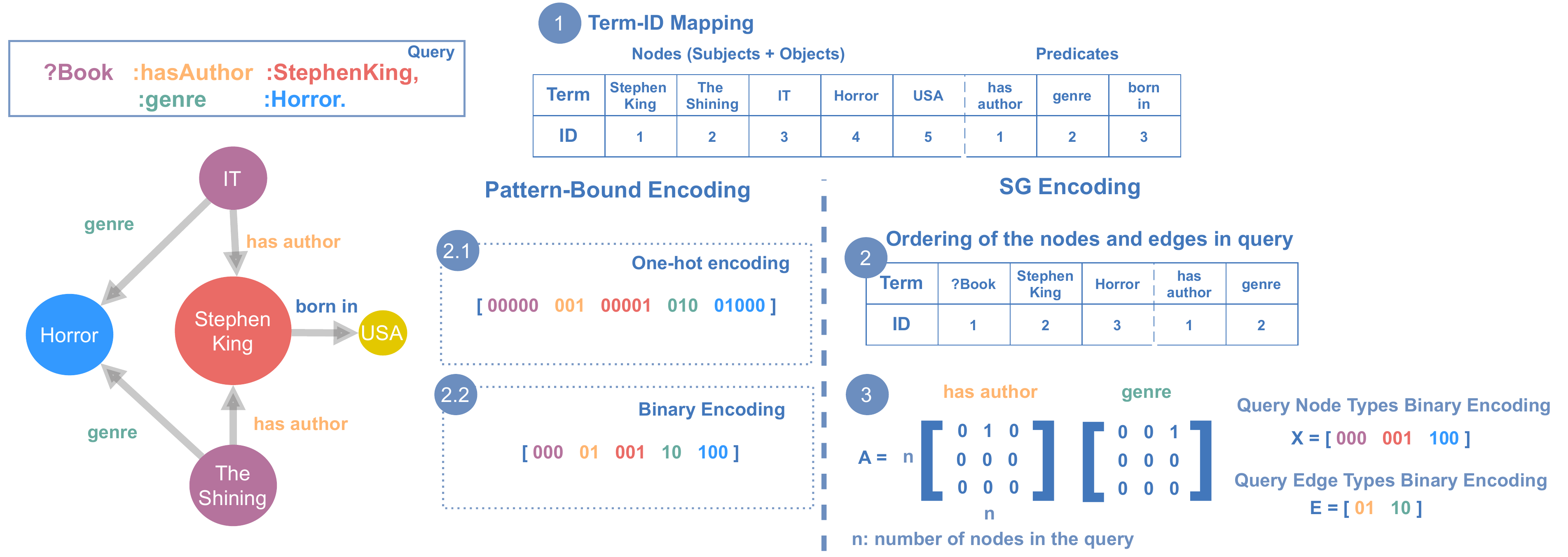}

  \caption{Encoding example}

  \label{fig:encoding-example}

\end{figure*}

Since LMKG represents the starting point and the first attempt for learning cardinalities of knowledge graph patterns, the focus is limited 
only on equality, i.e., presence or absence of terms.  
For cardinality estimation of range queries, one could modify the input encoding with histogram selectivity values,
which we will address in future work.

\section{Queries and Featurization}
\label{sec:join-queries}
A SPARQL query consists of triple patterns that can have unbound variables instead of one or more terms. 
The queries are distinguished based on their topologies in different classes: chain, star, tree, cycle, clique, petal, flower, and graph.
Although LMKG can be easily adapted for each of the query types, we focus on the two most common types of queries: star and chain~\cite{DBLP:journals/pvldb/BonifatiMT17}. Consider the following example:

\begin{verbatim}
SELECT ?x WHERE
  {  ?x  :hasAuthor  :StephenKing;
         :genre  :Horror. }
\end{verbatim}

The query pattern asks for all subjects (i.e., books) that are of genre Horror and have as author StephenKing.
This pattern corresponds to the class of star-shaped queries, in which the triples are centered around a single entity, i.e., same subject or object.
Formally, a subject star-shaped query consists of several triples $[(s_1,p_1,o_1),\ldots,(s_1,p_k,o_k)]$ such that all triples are centered around the same subject $s_1$.
A chain-shaped query joins triples, such that the object of the preceding triple is the subject of the next one. More specifically, a chain query consists of $k$ triples $[(s_1,p_1,o_1), (s_2,p_2,o_2),\ldots,(s_k,p_k,o_k)]$, such that $s_{i}=o_{i-1}$, where $i \in [2,k]$.
As an example, consider: 

\begin{verbatim}
SELECT ?x, ?y WHERE
  {  ?x  :hasAuthor  ?y.
     ?y  :bornIn  :USA. }
\end{verbatim}

The triples share a common unbound variable $?y$ which in the first triple has the role of an object and in the second of a subject. In this case, the solution will consist of all of the authors that have written a piece and are born in the USA.

We next delineate the single triple pattern encodings that constitute the more involved join query encodings. To model a triple pattern, we convert the triple terms into numerical values, each having an identifier in the range of 1 to the maximal number of nodes 
or predicates. Each of the terms in the triple is separately encoded. An encoding of a triple pattern is a concatenation of the encodings of the terms.
{\color{brown}
To capture term correlation, we always include all terms present in a subgraph and focus on a more compact representation of them.}
Since a simple integer encoding requires an ordinal relationship which is not present in categorical variables, as are the ones used in our case, in LMKG, we currently support two types of encodings for the triple patterns:

\begin{itemize}[leftmargin=*]
\item \textbf{One-hot encoding:} an encoding in which the bound terms involved in the triple are set to $1$. For example, if we assume that the total number of subjects in our knowledge graph is $3$, then the one-hot encoding for the subject with id $2$ will be $[010]$. 
An unbound term is treated as an absent one by setting its position in the encoding to $0$. 
A single triple encoding constructed out of one-hot encoded terms will occupy $O(|S|+|P|+|O|)$ space.

\item \textbf{Binary encoding:} an encoding in which the value of the term is represented with a binary digit. As an example, for a KG with $3$ unique subjects, the binary encoding of the subject with id $2$ will be $[10]$. 
As in the one-hot encoding, an absent term is simply encoded with a value of $0$.
The space of the binary triple encoding is now $O(log_2|S| + log_2|P| + log_2|O|)$. Although the binary encoding loses some of the expressiveness attained by the one-hot encoding and introduces larger complexity in the input, it is the preferred choice for encoding triple patterns. This is mainly because the knowledge graphs usually include data from various domains and consequently have a large number of unique subjects, predicates, and objects. Hence, a binary encoding will result in a smaller input dimensionality, making LMKG capable of dealing with heterogeneous knowledge graphs.
\end{itemize}

\subsection{Query Encodings}
\label{star-vectorization}

{\color{brown}
The most popular graph representations are Adjacency List, a list of nodes each pointing to a list of edges originating from the node, or an Adjacency Matrix $A$, where $A_{i,j,l}=1$ if node $i$ has an edge $l$ to node $j$. 
Considering space consumption, adjacency lists are preferred when graphs are sparse, as they do not represent absent edges. However, as  graphs get denser, the benefit of adjacency lists vanishes and can even lead to a higher space complexity compared to adjacency matrices.
Having the baselines in mind, we next present two possible directions for representing KGs subpatterns as well as outline the differences with the existing approaches:
}

\noindent
\subsubsection{Graph-based encoding} \label{encoding-strategy-our}

{\color{brown}The graph-based encoding should have the ability to represent every possible subgraph pattern of a specific size. Since the input of a neural network is of fixed size, the encoding should be fixed to the maximal number of features used to represent a dense subgraph of the KG.}
Knowing that the number of graph nodes (i.e., subjects and objects) is $d$ and the number of predicates is $b$, a KG encoding can be an adjacency matrix $A$ of space $O(d*d*b)$.
For the representation of a small subgraph pattern composed of several nodes and predicates, the complete matrix will be required, leading to huge matrices for real-life KGs.
Incontrovertibly, this severely impacts the training time as well as the accuracy of the model. 
{\color{brown}
Even for adjacency lists, we would need to encode the complete space of possibilities, leading to a dense graph representation, which, as already discussed, is not preferred over an adjacency matrix.
}

If we consider subgraph patterns that represent a subset of the complete graph, space consumption can be drastically reduced.
Following existing work on molecular graphs~\cite{DBLP:journals/corr/abs-1905-13372,DBLP:conf/iclr/ShiXZZZT20}, we can represent a KG subgraph with $n$ nodes as $G = (A, X)$, where $A \in \{0,1\}^{n*n*b}$ and $X \in \{0,1\}^{n*d}$. 
Although suitable for molecular graphs where the number of distinct relations is small, in most KGs this encoding creates a large sparse matrix $A$, due to the high number of relations.

Following this line of thought, to circumvent large matrices created from the existing representations and adapt them to knowledge subgraphs, LMKG employs a novel subgraph encoding, termed \textit{SG-Encoding}. 
We define a subgraph pattern to have $n$ number of nodes and $e$ number of predicates, where $n<d$ and $e<b$.
A subgraph pattern is represented as $SG = (A, X, E)$, where $A$ is the adjacency tensor, $X$ is the node feature matrix and $E$ is the predicate feature matrix.
Given an ordering of the nodes and predicates from the subgraph pattern, we define $A \in \{0,1\}^{n*n*e}$, $X \in \{0,1\}^{n*d}$ and $E \in \{0,1\}^{e*b}$, where $A_{ijl}=1$ if there exists an edge $l$ between the $i$-th and $j$-th node.
We set $X_{im}$ to $1$ if the $i$-th node is of type $m$ and $E_{lk}$ to $1$ if the $l$-th predicate is of type $k$. 
Intuitively, if $b$ is smaller than a threshold value, we can eliminate the matrix $E$, 
as shown in previous work~\cite{DBLP:conf/iclr/ShiXZZZT20}.

Initially, each row in $X$ and $E$ is a one-hot encoding of size $d$ and $b$, respectively, where the rows in $X$ represent the node type and the rows in $E$ the edge type. For a more compact representation, we provide a further modification of the matrices $X$ and $E$, where instead of one-hot encoding, we employ a binary encoding for each of the nodes and edges in the subgraph.
More specifically, the encoding is modified such that $X \in \{0,1\}^{n*\lceil{log_2|d| + 1}\rceil}$ and $E \in \{0,1\}^{b*\lceil{log_2|e| + 1}\rceil}$, thus, drastically reducing the space for the subgraph representation. 
Although not all subjects are objects and vice versa, when encoding SPARQL queries, there is still a need to express that they can be equal. This is especially important for chain queries, where the subject of one triple pattern is the object of another. Consequently, there is only a single node matrix and not two separate ones. 

{\color{brown}
The SG-Encoding can handle composite queries as it can simultaneously represent more than one type of a query. To initiate the SG-Encoding, an ordering of the nodes and edges is required. }
The subgraph encoding for an example star query is illustrated in \figurename~\ref{fig:encoding-example} (right). We show the encoding of a query by breaking down the process into three main steps. In Step 1, executed in the creation phase (Section \ref{trainign-data-creation}), a mapping of the nodes and predicates to an integer id is created. 
Next, SG-Encoding creates an ordering of the nodes and predicates for the given query which is illustrated in Step 2. Finally, for a predefined $n=3$ and $e = 2$, the parts of the encoding are created. For the subpattern $?Book\ :\!hasAuthor\ :\!StephenKing.$, we set $A_{001} = 1$, indicating that $:\!hasAuthor$ is the first predicate in the edge order, the unbound variable is the first and $:\!StephenKing$ is the second in the respective node order. The mapping of nodes and edges in the query to their ids is represented by $X$ and $E$, respectively.  
From the example, the first two bits in $E$ indicate that the predicate, $:\!hasAuthor$, has id 1.

\subsubsection{Pattern-bound encoding} represents an encoding that works only for a model tailored for a certain type of query. 
For subject-oriented star patterns, the pattern-bound encoding requires ordering of the predicate-object pairs connected to the subject.
More specifically, if we have a triple set of size $t$ that is centered around a single unbound subject, the star encoding is a concatenation of all of the $t$ pair encodings and the subject encoding. The terms can be represented either by a one-hot encoding or a binary encoding. An example encoding for a star query is shown in  \figurename~\ref{fig:encoding-example} (left). In the first step, a mapping of the nodes and the predicates to an integer id is created. Then every term in the query is encoded using either the one-hot or the binary encoding, resulting in the encodings shown in Step 2.1 and Step 2.2, respectively.

Unlike in star-shaped queries, in chain-shaped queries the ordering of the nodes and edges is evident. Intuitively, a predicate 
connected to a subject is its descendant and its position in the order will follow the one of the subject. Similarly the same holds for the object which has as an antecedent the predicate. Given an ordering of the nodes and edges, we can encode the chain query as a concatenation of the encodings of the separate terms, encoded either with a one-hot encoding or preferably for larger patterns, a binary encoding.

{\color{brown}
As previously discussed, an adjacency list representation is preferable when the subgraph query pattern is sparse. This is the case when the subgraph query pattern is fixed since we limit the capability to encode only the specific subgraph pattern and not all possible variants, i.e. a sparse representation. 
As one can observe, a serialization of an adjacency list resembles the pattern-bound encoding. More specifically, a flattened adjacency list for a star graph pattern will directly correspond to the star encoding. However, for a chain query, an adjacency list will be larger than the pattern-bound encoding since by knowing that an object in a triple will be a subject in the next one, we further remove redundant nodes.
}

LMKG can apply both encodings depending on the selected query types that are to be learned by the models. If a model is chosen for learning only a specific type of query, then a pattern-bound approach is sufficient.
Although by using the matrices $X$ and $E$ without $A$ from the SG-encoding will lead to the pattern-bound approach, the matrix $A$ is crucial for representing different types of query topologies or combinations of them in a single model. For instance, the same model may later be trained on tree or clique queries of a predefined size.

\section{Neural network specification}
\label{sec:lmkgUsedModels}
{\color{brown}
Neural networks are capable of detecting patterns in high dimensional data, which is highly useful when it comes to the high number of terms co-occurring in a KG.}
Following previous work on learned cardinality estimation in relational databases, e.g., ~\cite{MSCN,SimpleModels,LocalDeepLearningModels,Naru}, LMKG utilizes two models, a supervised and an unsupervised model, i.e.,  LMKG-S and LMKG-U.

\subsection{Supervised Model (LMKG-S)}

\begin{figure}[!t]

 \center
  \includegraphics[width=0.47\textwidth]{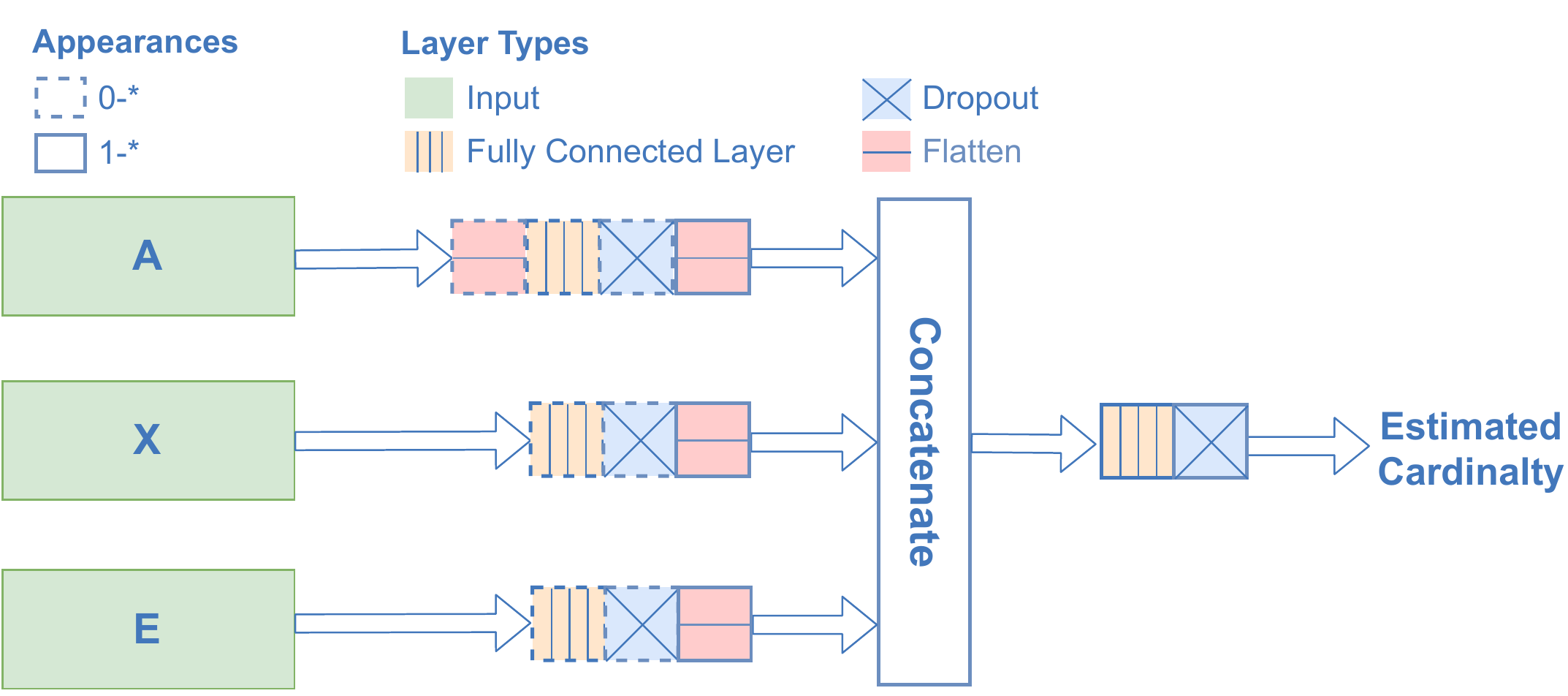}
  \caption{LMKG-S architecture}
  \label{fig:figure-lmkg-s-architecture}

\end{figure}

Supervised deep learning models efficiently approximate non-linear functions, thereby providing an estimated output for a given input. By tuning the number of layers and neurons, that means, increasing the set of learned parameters, these models can fit different levels of non-linearity and learn different patterns from the data. To benefit from the expressiveness of deep learning models, in LMKG, we first address cardinality estimation as a supervised learning problem.
A standard way of modeling the supervised regression task is with the usage of multi-layer perceptions. The network consists of an input layer, an arbitrary number of hidden layers, and an output layer. The number of neurons and layers can vary and are set according to the complexity of the input. In the case of LMKG, the structure of the neural network changes according to the encoding of the input, tailored to the considered query types, i.e., star and chain queries. During the training process, the neural network optimizes itself based on the provided example pattern queries with precalculated cardinalities.
The architecture of LMKG-S is shown in  \figurename~\ref{fig:figure-lmkg-s-architecture}. Initially, the cardinalities are log scaled followed by a min-max scaling. Depending on the size of the input, some of the layers can be optional, displayed with a dashed border in  \figurename~\ref{fig:figure-lmkg-s-architecture}. Nevertheless, X, A, and E are always concatenated once flattened and propagated through one or multiple layers. Every fully connected layer except the output layer uses ReLU, a non-linear activation function defined as $f (x) = max (0, x)$. The output layer uses a sigmoid function $f(x) =\dfrac{1}{1 + e^{-x}} $, having an output between $0$ and $1$ that is suitable for the already scaled values. After the exploration of different objective functions, we have concluded that an adequate loss function for cardinality is the mean q-error. The q-error is defined as the relation between the true cardinality and the estimate, i.e., $q\_error(y, \hat{y}) = max(\dfrac{\hat{y}}{y},\dfrac{y}{\hat{y}})$. When training for one specific type of query, LMKG-S can also employ the pattern-bound encoding. This encoding is then simply a flat vector composed of its terms and can be directly provided as an input to the neural network. 

Although different approaches exist for handling graph data, the usage of lightweight models results in faster execution and lower memory consumption. 
    
\label{ssed:supervised-manner}

\subsection{Unsupervised Model (LMKG-U)}
\label{explanation:unsupervised-manner}
A deep autoregressive model is an unsupervised deep learning model that efficiently estimates a joint distribution $P(x)$ from a set of samples.
An autoregressive property dictates that, for a predefined variable ordering, the output of the model contains the density for each variable conditioned on the values of all preceding variables~\cite{Goodfellow-et-al-2016}.
Therefore, given as input $x = [x_1, x_2, ..., x_d]$, the autoregressive model produces $d$ conditionals which, when multiplied, will result in the point density $P(x)$. 
Formally, for $x = [x_1, x_2, ..., x_d]$:
$$P(x)\!=\!P(x_1)P(x_2|x_1)...P(x_d|x_1, ... , x_{d-1})\!=\!\prod_{d=1}^{D} P(x_d|x_{<d})$$

Recently, the use of deep autoregressive models has been proposed as another way of estimating the query selectivities in relational databases~\cite{MultiAttributeCE, Naru}. Following previous work, we introduce LMKG-U where we adapt autoregressive models for cardinality estimation in KGs. The autoregressive models summarize relations in an unsupervised manner, resulting in a synopsis that is ideal for representing KG characteristics. 
In the recent years, several autoregressive architectures have been proposed~\cite{DBLP:conf/icml/DurkanN19,DBLP:conf/icml/GermainGML15}. As one of them, MADE~\cite{DBLP:conf/icml/GermainGML15} can maintain the autoregressive property by adequately masking particular weights in the layers. In our work, we use ResMADE, a modified version of MADE enhanced by residual connections.

In autoregressive models, the cardinality estimation for a given query is calculated by using the conditional probabilities of the terms in the query.
Thus, for a query involving $k$ triples $qp_k = n_1, p_1, n_2, p_2, ... , p_{k}, n_{k+1}$, the density can be estimated as $P(qp_k) = P(n_1)P(p_1|n_1)...P(n_{k+1}|n_1,p_1, ... , n_{k}, p_{k})$. For a chain-shaped query, $n_i$ is an object in the triple $t_{i-1}$ and a subject in the triple $t_i$ where $i \in [2,k]$. For a star-shaped query, $n_1$ is the only subject and $n_i$ where $i \in [2, k+1]$ are objects connected to the subject through $p_j$ where $j \in [1,k]$.
To avoid generating nonrelevant samples for the query terms by using rejection sampling, as in previous work~\cite{DBLP:books/daglib/0023091, Naru}, we employ likelihood-weighted forward sampling.
Similar to Bayesian Networks, terms are sampled in a predefined order and are impacted by the previously created terms, in the end constructing a weighted particle. The weights incorporate the likelihood of the already generated terms accumulated throughout the sampling. Thus, the new term value is impacted by the probability of the already generated bound term values.

Although we advocate the usage of two different types of encodings, the SG-Encoding can contribute to increased complexity of the input. For datasets with a high number of unique terms, SG-Encoding will render LMKG-U ineffective. Hence, the preferred encoding when using LMKG-U is the pattern-bound encoding. 
Still, since the KGs are highly heterogeneous, the usage of one-hot and binary encoding may lead to large models. Therefore, to further reduce the model size, we apply an embedding on each of the terms in the pattern-bound encoding. The embedding will capture the term similarities and drastically reduce the size of the model but it may result in lower accuracy. 

During the exploration of possible models for KG cardinality estimation, we also examined graph-based autoregressive flows~\cite{DBLP:conf/iclr/ShiXZZZT20}. However, unlike in the autoregressive model, the conditional probabilities are not tractable and thus, we cannot directly use these models for our use-case.

\section{Approach and Models Overview}
\label{sec:appr_mode_overview_new}
\subsection{Training Data Creation}
\label{ssec:training_data_creation_appr_overview}
The endless combinations of queries emerging from the directed graph structure directly impact the training data creation as well as, the accuracy of the cardinality estimation model. 
For smaller queries and homogeneous KGs, the complete set of patterns for a specific size can be created. However, as the size increases, the creation of all the possible patterns of size $k$ creates a combinatorial problem. Therefore, to generate the training data, proper sampling has to be conducted. 
The sampling has to satisfy the scaled-down property of the KG, meaning that the samples should possess similar properties as the original KG~\cite{DBLP:conf/kdd/LeskovecF06}. As Leskovec and Faloutsos~\cite{DBLP:conf/kdd/LeskovecF06} show, the best performance for a scaled-down sampling is achieved by the random walk (RW) sampling since it is biased towards highly connected nodes. Furthermore, RW preserves the property even when the sample size gets smaller. To simulate RW for star patterns of size $k$, we randomly pick a starting node and then simulate a random step $k$ times from the starting node. Similarly, for chain patterns, we start a random walk from a randomly selected node and stop once the required size is reached. Although we use RW sampling for generating training data, efficient sampling in KGs is still a challenging area and as further shown, the main cause of inaccurate model estimation is the quality of the samples, especially visible in KGs with many unique terms.

{\color{brown}\subsection{Grouping of Models and Models' Analysis}
\label{ssec:models_grouping_and_analysis}
To try to deal with virtually unlimited combinations of queries, LMKG can create compounds of the models grouped by different criteria, allowing the choice between:

\noindent\textbf{Single learned model} that can be used for the complete knowledge graph and all types of queries, including star-shaped and chain-shaped queries with $k$ number of joins.

\noindent\textbf{Query type grouping} that creates separate models each specialized for different types of queries. Each model is tailored to a query type and it can use both encodings.

\noindent\textbf{Query size grouping} that creates a single model for a range of queries, grouped by their size, e.g., a single model can be created for patterns up to size 4 and another for larger ones.

LMKG offers the creation of different estimators characterized by the specific encoding, the type of learning, and grouping. We next delineate the benefits and the downsides of the models along these dimensions and how they address the challenges in cardinality estimation in KGs:
 
\noindent\textbf{Encoding:}
Both encodings are capable of addressing two of the challenges. They capture the term intercorrelations for a specific query, by not leaving out any of them in the encoding. Additionally, both encodings can express many self-joins, assuming that one-hot term encoding is not used.

 \textit{Pattern-bound encoding:}
 simple to implement and has a small dimensionality since it is tuned to a specific query. However, when the patterns consist of reoccurring nodes or predicates, in some cases they are repeated in the encoding, which may lead to higher dimensionality. The encoding is not generalizable to different query topologies, thus, it requires the creation of multiple models and needs higher maintenance.

 \textit{SG-Encoding:} unlike the pattern-bound encoding, addresses the third challenge since it can simultaneously represent different query topologies. On the other hand, it can have a larger dimensionality which is especially noticeable when all the terms in the query are distinct.

\noindent\textbf{Supervision:}
Both models capture term correlations but are highly impacted by the training data sample quality. LMKG-S needs to capture enough representative queries which describe the common workload. LMKG-U samples have to have the same ratio as the original data which can be challenging for larger dimensions.

\textit{LMKG-S} has a faster training and prediction time, as well as a smaller memory footprint. However, the training data creation is more time consuming since unbound variables need to be included. Additionally, generalizing to queries that are far from the training data is challenging and somewhat impacted by the slight overfit for better results, leading also to worse estimates for outliers.

\textit{LMKG-U} can create training data faster since the model learns only from bound terms.
It also captures the term intercorrelations better than LMKG-S, producing highly accurate cardinality estimates especially suitable for skewed datasets.
However, LMKG-U has a larger memory footprint and higher training and prediction time than LMKG-S. This is especially noticeable when working with heterogeneous datasets that have a high number of unique terms.

\noindent\textbf{Grouping of Models:} The grouping mainly affects the accuracy and the creation time of the models.

\textit{Single learned model:} suitable for small memory budgets and homogeneous and smaller KGs. It requires less tuning and maintenance during the run-time phase. However, it may produce lower accuracy for heterogeneous KGs due to the high number of patterns that affect the quality of the samples.

\textit{Query Type or Query Size Grouping:} enables parallel creation of the training data and the models, leading to an excessive time reduction. However, having multiple models may lead to increased memory consumption
and maintenance. 

Combining the models depends on the overall model creation budget as well as the dataset characteristics.
If a smaller model creation time is needed, LMKG-S is preferred over LMKG-U. If the training data is smaller, LMKG-U can still be considered, even in combination with LMKG-S.
If there are no training time constraints, then the data characteristics should be examined. For instance, for star-queries over datasets having only several nodes with a huge number of in- or out-
degree i.e., skewed distribution, LMKG-U is preferred. However, when many rare terms appear and we are working with chain queries, the training data sampling may be worse for LMKG-U. Thus, in a situation where these two cases appear, a combination of both LMKG-U and LMKG-S may be the preferred approach. 
A single compound incorporating a supervised and an unsupervised model, as one model, for estimating a single query cardinality is currently out of the scope of this paper and left for future work.
}

\pdfoutput=1

\section{Experiments}
\label{sec:experiments}

We next present the results from the experiments, conducted on both synthetic and real-world datasets,
organized into two main blocks. The first block incorporates a thorough investigation of the LMKG models. 
The second block comprises a comparison of the LMKG models with state-of-the-art cardinality estimation approaches for KGs. 

\noindent\textbf{Setup:}
We have implemented the LMKG models in TensorFlow and carried out the experiments for the learned models on an NVidia GeForce RTX 2080 Ti GPU. We have evaluated the competing approaches on a server with an Intel Xeon E7-4830 v3 CPU @ 2.10GHz and 1 TB RAM. For all the competitive approaches except for the Characteristics Sets (CSET) approach, we have used the publicly available implementation provided by the authors of the G-CARE framework \cite{gCare}. For the CSET approach, we followed the reference paper and tried to implement the presented algorithm to the best of our capabilities. {\color{brown}We decided to reimplement CSET ourselves as the G-CARE adaptation for chain queries returned unrealistically high estimates that negatively affect the final results.} 
Like in the G-CARE framework, the sampling approaches are run 30 times and the resulting cardinality estimate and execution time are an average over the 30 samples. The timeout of the methods is set to 5 minutes.

In Table~\ref{tab:datasetExperimentsSpecification} we show the specifications for the experiments. 
As mentioned, LMKG currently supports chain and star queries as the most typical query types~\cite{DBLP:journals/pvldb/BonifatiMT17}.
To see the models' performance, we group the queries into buckets depending on the query result size, with boundaries defined by log with base 5. For observing the performance for different query complexities we choose a query size of 2, 3, 5, and 8, for both star and chain queries, including at least 1 unbound variable.

\begin{table}
\caption{Experiment and dataset specifications}
\label{table:datasets-explanation}
\scalebox{0.8}{
 \begin{tabular}{|c | c |}
 \hline
    Dataset & SWDF, LUBM20, YAGO \\
    \hline
    Topology & Chain, Star \\
    \hline
    Result Size & $[5^0,5^1], [5^1,5^2], ..., $ \\
    \hline
    Query Size & 2, 3, 5, 8 \\
    \hline
\end{tabular}
\quad

 \begin{tabular}{|c | c | c| c|}
 \hline
    Dataset & SWDF & LUBM20 & YAGO \\
    \hline
    Triples & \texttildelow 250K & \texttildelow 2.7M & \texttildelow 15M \\
    \hline
    Entities & \texttildelow 76K & 663K & 12M \\
    \hline
    Predicates & 171 & 19 & 91 \\
    \hline
\end{tabular}
}

\label{tab:datasetExperimentsSpecification}
\end{table}

\begin{figure}[!t]
  \begin{minipage}{0.45\linewidth}
    \includegraphics[width=\linewidth]{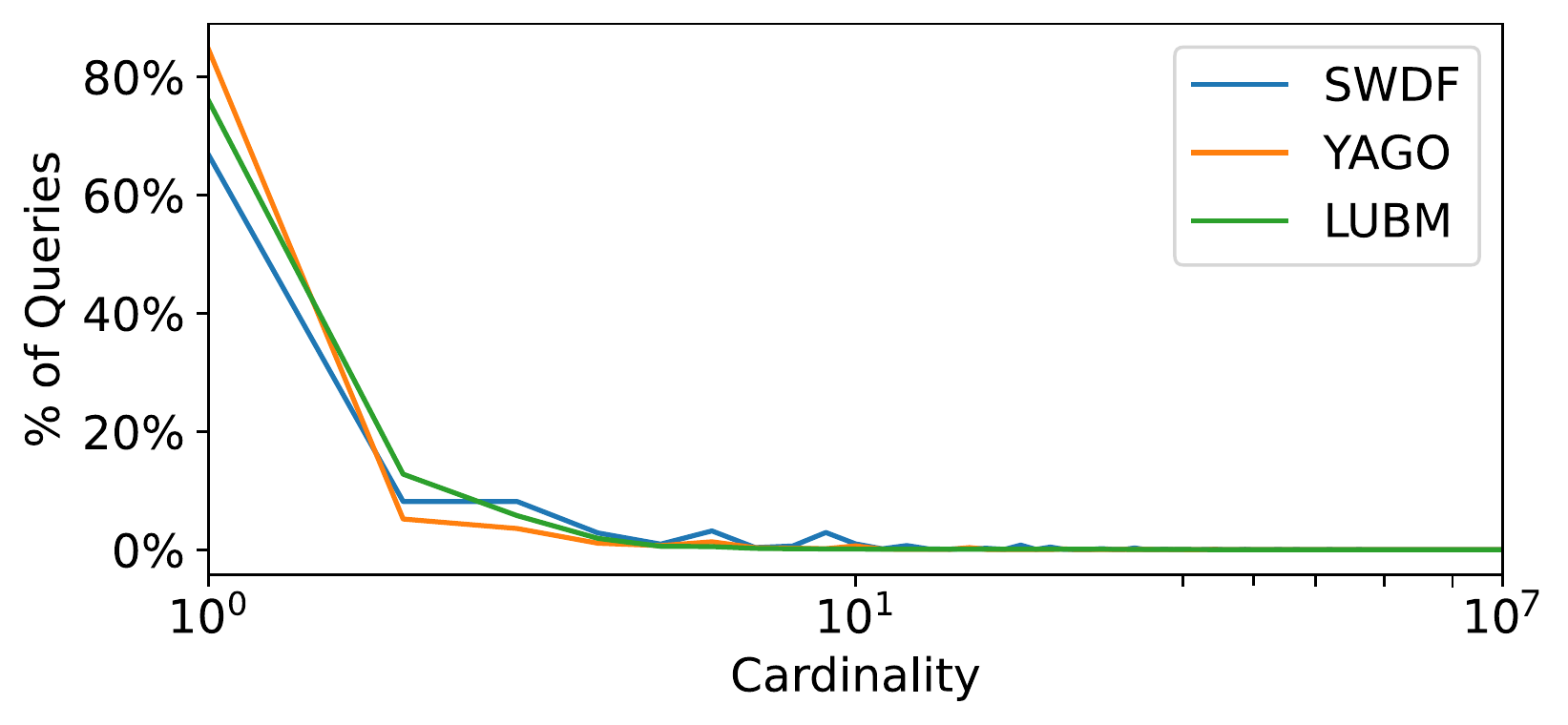}
    \centering
    \captionof{figure}{\textcolor{brown}{Datasets distribution}}
    \label{fig:datasets_distribution}
  \end{minipage}
\hfill
\begin{minipage}{0.45\linewidth}
    \includegraphics[width=\linewidth]{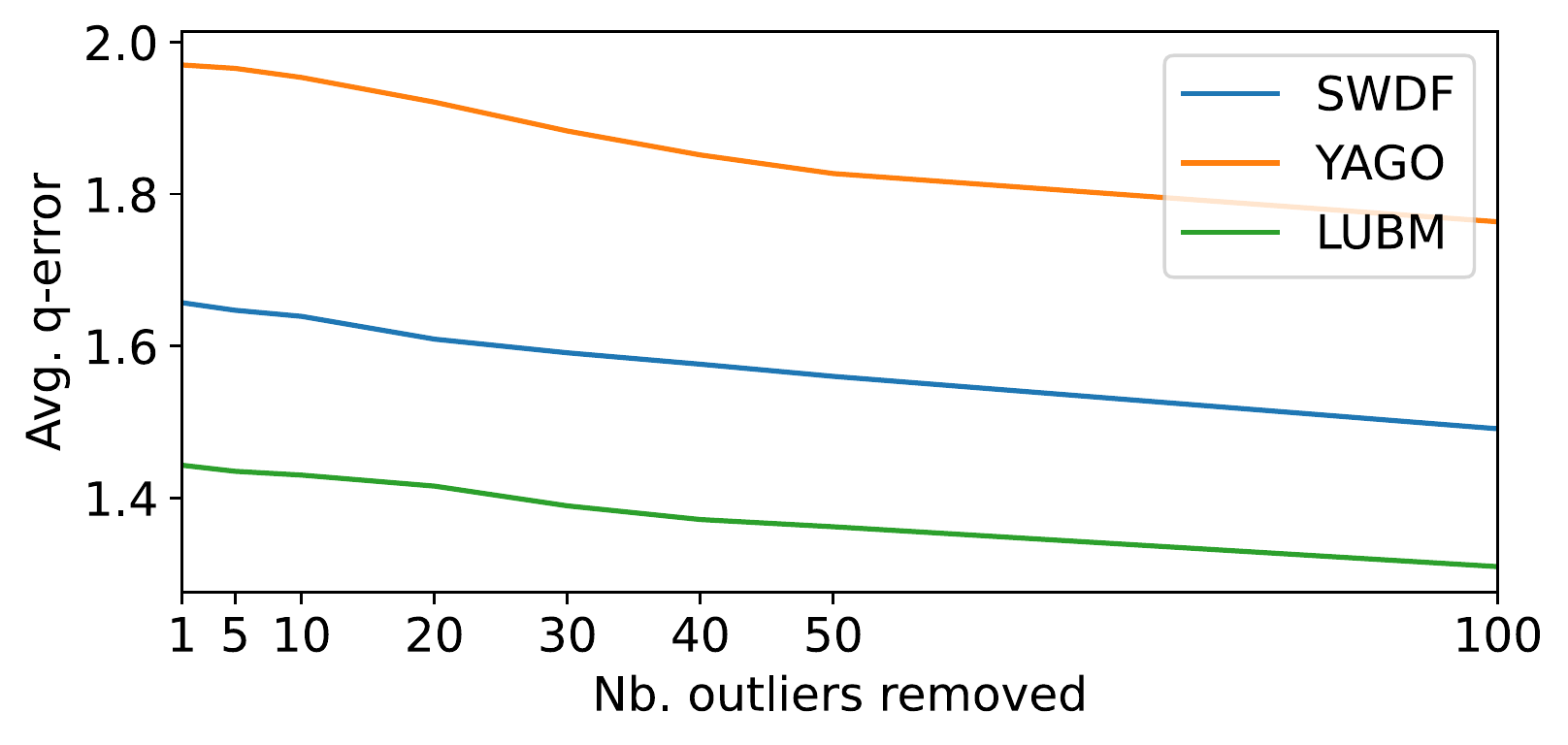}
    \centering
    \captionof{figure}{\textcolor{brown}{Impact of outliers}}
    \label{fig:outlierImpact}
  \end{minipage}
\vspace*{-6mm}
\end{figure}

\noindent\textbf{Datasets:}
We use one synthetic and two real-world datasets with different characteristics (shown in Table~\ref{tab:datasetExperimentsSpecification}). We use the Semantic Web Dog Food (SWDF)~\cite{SWDF} which although contains a smaller number of triples,  has a high number of interconnections between the terms. We use LUBM~\cite{LUBM}, a widely used RDF benchmark, for which we use a scaling factor $20$. We also use YAGO~\cite{YAGO} as a larger knowledge base, chosen for its large number of distinct term values. 
{\color{brown}
The data analysis shows that the query cardinality follows a skewed distribution. As depicted in \figurename~\ref{fig:datasets_distribution}, where we show the query cardinality for the datasets by averaging over the different query sizes, the vast amount of queries have a small cardinality. Moreover, there is a presence of extremely large query cardinalities, i.e., outliers, which highly impact the accuracy of the models.}
For training the models we create training data according to the explanation in Section~\ref{ssec:training_data_creation_appr_overview} where the sample size depends on the dataset characteristics.  

\noindent\textbf{Competitors:}
As competitors, we use the approaches in the recently developed benchmarking framework G-CARE~\cite{gCare} and an adaptation of the supervised estimator \textcolor{brown}{MSCN}~\cite{MSCN}.
G-CARE includes sampling-based and summary-based techniques, some native to knowledge graphs and others adapted from relational databases. The competitors are:
\noindent\underline{Summary-based approaches:}
\noindent\textit{Characteristic set}~\cite{DBLP:conf/icde/NeumannM11} summarizes entities based on their emitting edges and it is specifically tailored for star-shaped queries. 
\noindent\textit{SUMRDF}~\cite{DBLP:conf/www/StefanoniMK18} is a graph summarization approach that estimates the cardinality by relying on the possible world semantics. 
\noindent\underline{Sampling-based approaches:}
\noindent\textit{WanderJoin}~\cite{DBLP:conf/sigmod/0001WYZ16}, 
performs random walks directly on top of the KG by considering each triple as a vertex and a join as an edge.
\noindent\textit{Impr}~\cite{DBLP:conf/icdm/ChenL16} 
uses random walks for estimating graphlet counts. 
\noindent\textit{Join Sampling with Upper Bounds (JSUB):}~\cite{DBLP:conf/sigmod/ZhaoC0HY18} random walk sampling approach for sampling over joins, adapted for producing estimates of the upper bound of the cardinality.
\noindent\textit{MSCN-n:} a multi-set convolutional network using query features as sets and $n$ materialized samples. In MSCN we perform self-joins over a single table to allow KG queries and always train on the same queries as LMKG-S. We use two variants, MSCN-0 and MSCN-1k with $0$ and $1000$ samples, respectively, with the same hyperparameters and trained until convergence. 
 
\noindent\textbf{Generation of Test Queries:}
For generating the test queries, we vary the query topology, the query size, and the query result size. 
For a specific query size and a fixed topology, we select 600 queries where each query is drawn from a bucket for a specific result size. Although we try to select the same number of queries from each bucket, the buckets including queries with a larger result size, i.e., cardinality, are usually smaller. It is also important to note that we limit the graph patterns to include only bounded predicates, due to the competitors' limitations to answer queries with unbounded ones.
\begin{figure}[!t]
\centering
\subfloat[LMKG-U (sample from LUBM)]{
    \centering
    \includegraphics[width=0.47\linewidth]{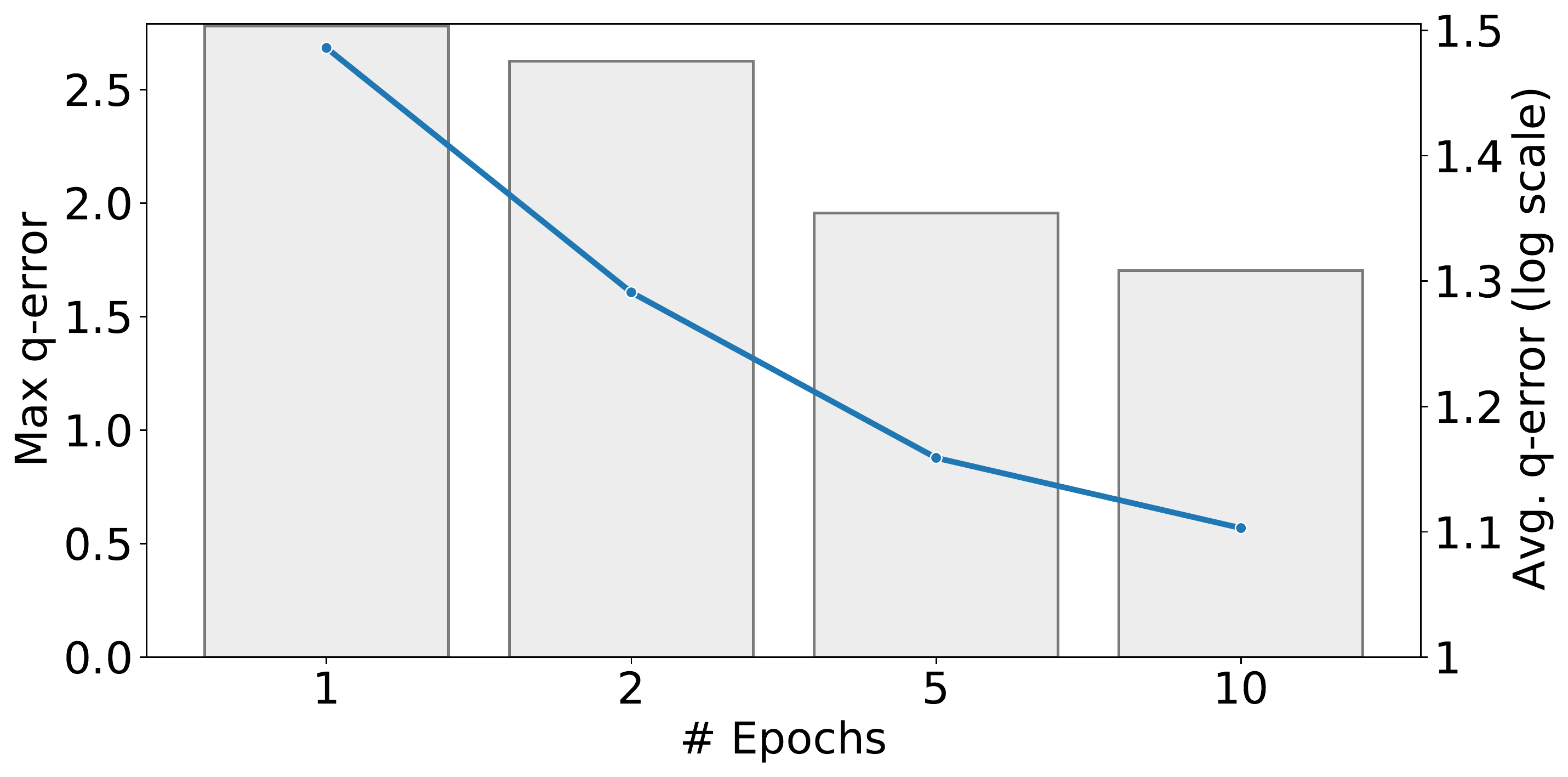}
    \label{fig:epochs-error-lmkg_u}
}%
\hfill
\subfloat[LMKG-S (sample from LUBM)]{
  \centering
  \includegraphics[width=0.47\linewidth]{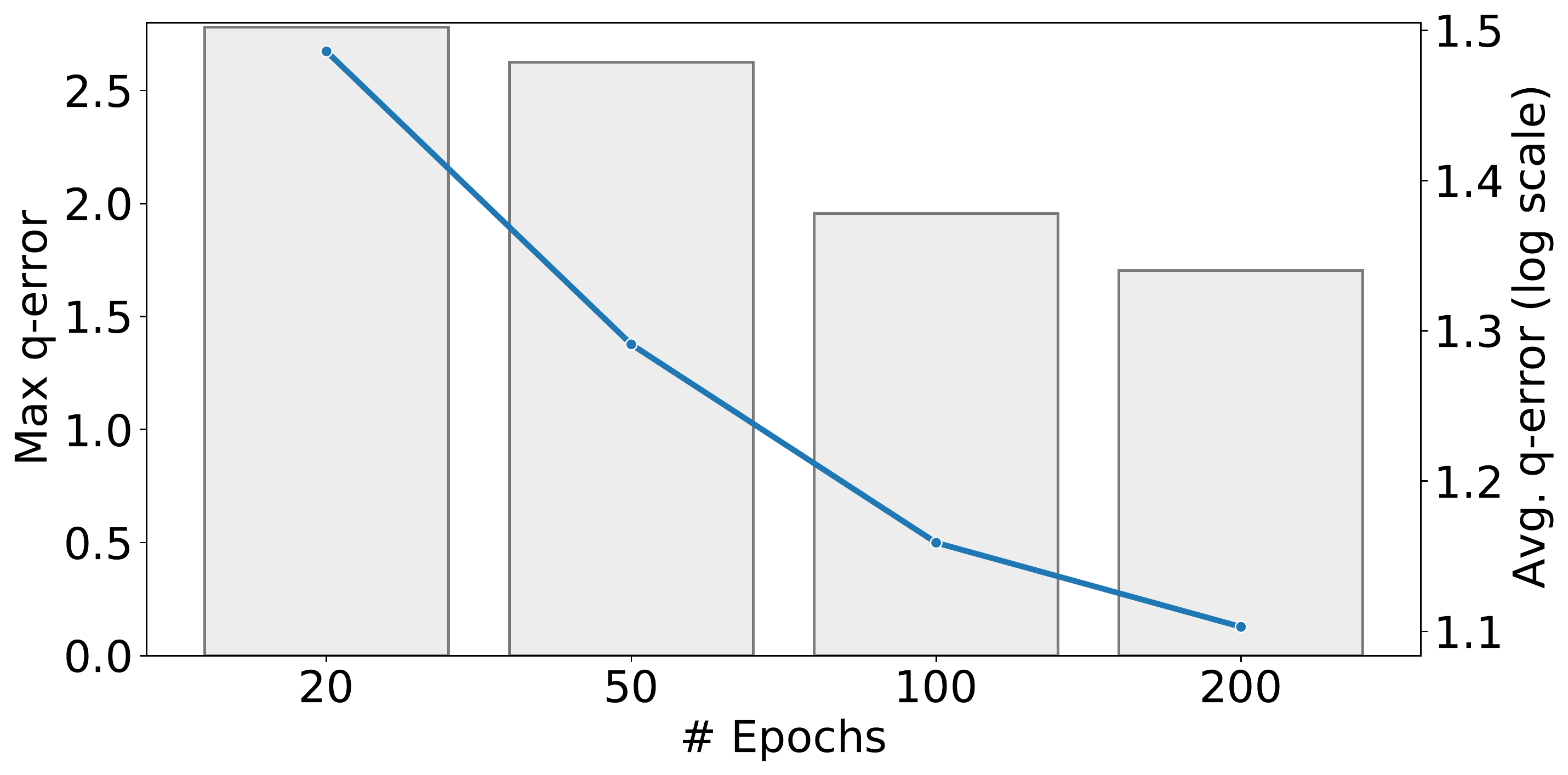}
  \label{fig:epochs-error-lmkg_s}
}
\caption{\textcolor{brown}{Training time vs. accuracy - bars show max q-error; dots show avg. q-error}}
\label{fig:epochs-error}
\end{figure}
It is important to point out that although LMKG-S can easily handle heterogeneous datasets with many distinct terms, LMKG-U can not. LMKG-U is an autoregressive model where the size of the model parameters is directly affected by the number of terms and their unique values. 
Hence, as the number of unique term values increases so will the model. This directly affects our experiments for the YAGO knowledge base. As previously depicted, YAGO contains a huge number of distinct term values (Table~\ref{tab:datasetExperimentsSpecification}). With the current setting, LMKG-U is not able to learn the complete set of queries of size 3 and beyond.
Since addressing this limitation is part of our ongoing work, we remove LMKG-U for the comparison with YAGO.

\subsection{Analysis of LMKG Models}
\label{ssec:comb-spec}
We next discuss some of the factors impacting LMKG:

\noindent
\textbf{Hyperparameter Tuning:}
We conducted experiments varying hyperparameters such as epochs, hidden units, and layers.
\figurename~\ref{fig:epochs-error} shows how two of the accuracy metrics change through the training process. The metrics are calculated after several epochs. 
After a reasonable number of epochs, the approaches reach satisfactory results for both the average and the maximal q-error.
For our experiments, we choose $200$ epochs for LMKS-S and $5$ epochs for LMKG-U to balance the training time and accuracy. 

{\color{brown}For LMKG-S, for SWDF, LUBM, and YAGO, on average, an epoch takes $13, 16,$ and $48$ seconds, respectively. The mentioned training times are directly affected by the sample size. Due to the higher complexity of the model in the presence of numerous terms, LMKG-U requires a longer training time. For the sample size that we have considered, for SWDF and LUBM, on average, an epoch takes $13$ and $18$ minutes, respectively. However, when we want to create a larger sample size, because of many unique terms and possible query patterns, an epoch in LMKG-U can take up to $42$ and $48$ minutes for queries with size 5 in LUBM and SWDF, respectively. 

We varied the number of hidden units ($256, 512, 1024$) and hidden layers (2--4) depending on the dataset characteristics. We found that $2$ or $3$ layers of $512$ neurons are often acceptable for both models. If not specified otherwise, we always report the combination producing the best results.}

\begin{figure}[!t]
\vspace*{-4mm}
\centering
\subfloat[Star queries]{
    \centering
    \includegraphics[width=0.47\linewidth]{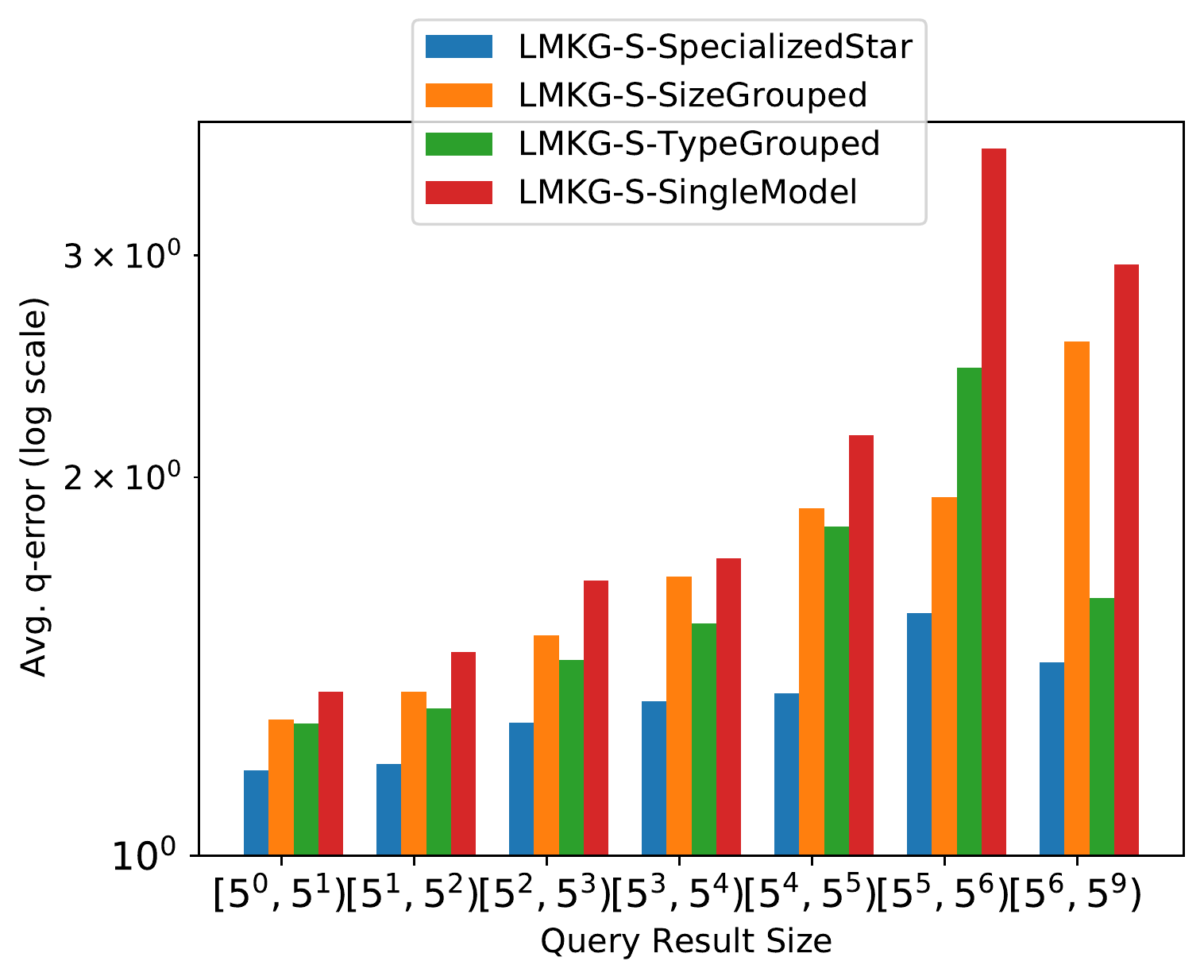}
    \label{fig:spec-vs-comb-star}
}%
\hfill
\subfloat[Chain queries]{
  \centering
  \includegraphics[width=0.47\linewidth]{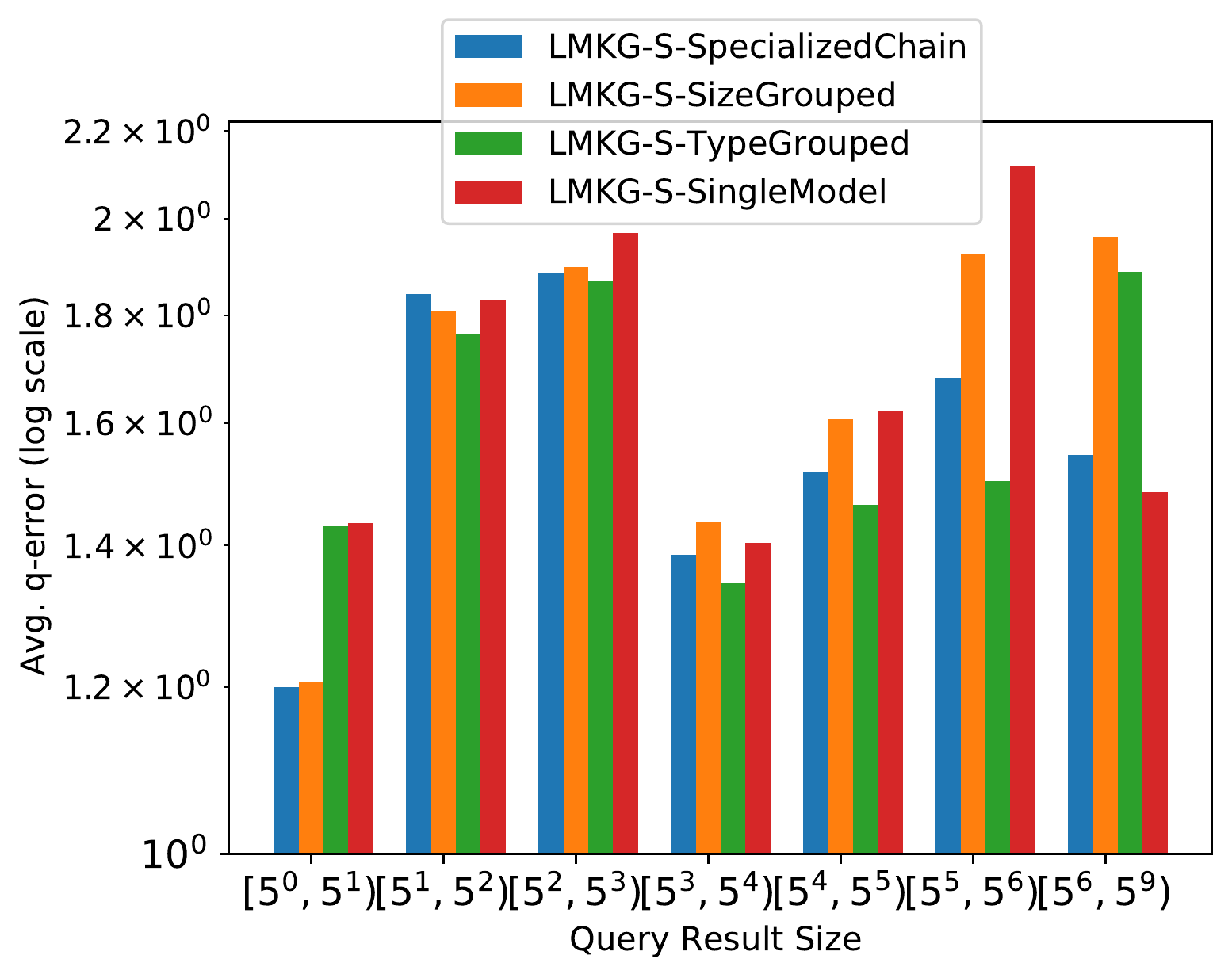}
  \label{fig:spec-vs-comb-chain}
}
\caption{\textcolor{brown}{Avg. q-error comparison between specialized and combined models}}
\label{fig:avg_q-err}
\end{figure}

\begin{figure}[!b]
\centering
\subfloat[SWDF]{
    \centering
    \includegraphics[width=0.47\linewidth]{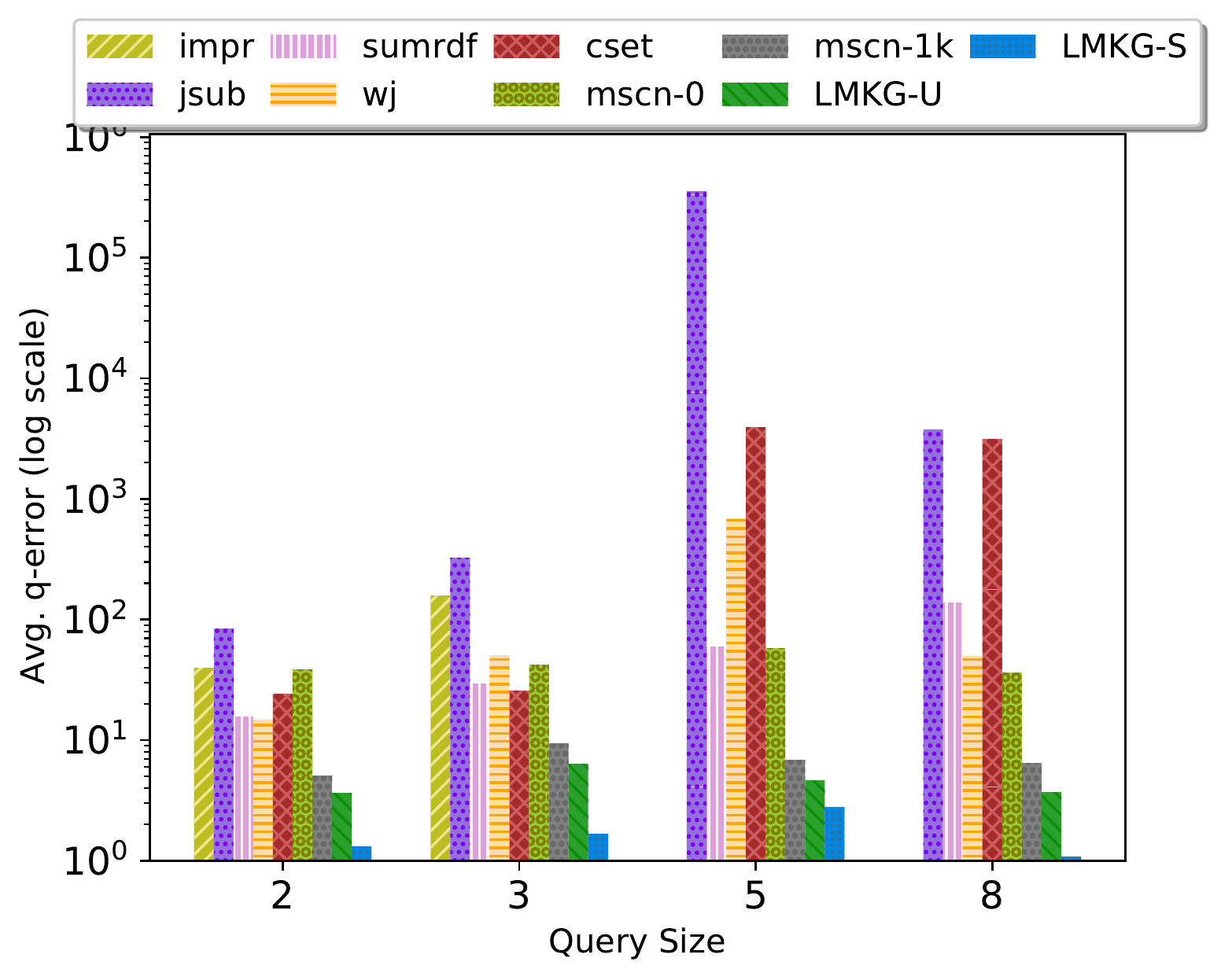}
    \label{fig:avg_q-err_query_size_swdf}
}%
\hfill
\subfloat[LUBM]{
    \centering
    \includegraphics[width=0.47\linewidth]{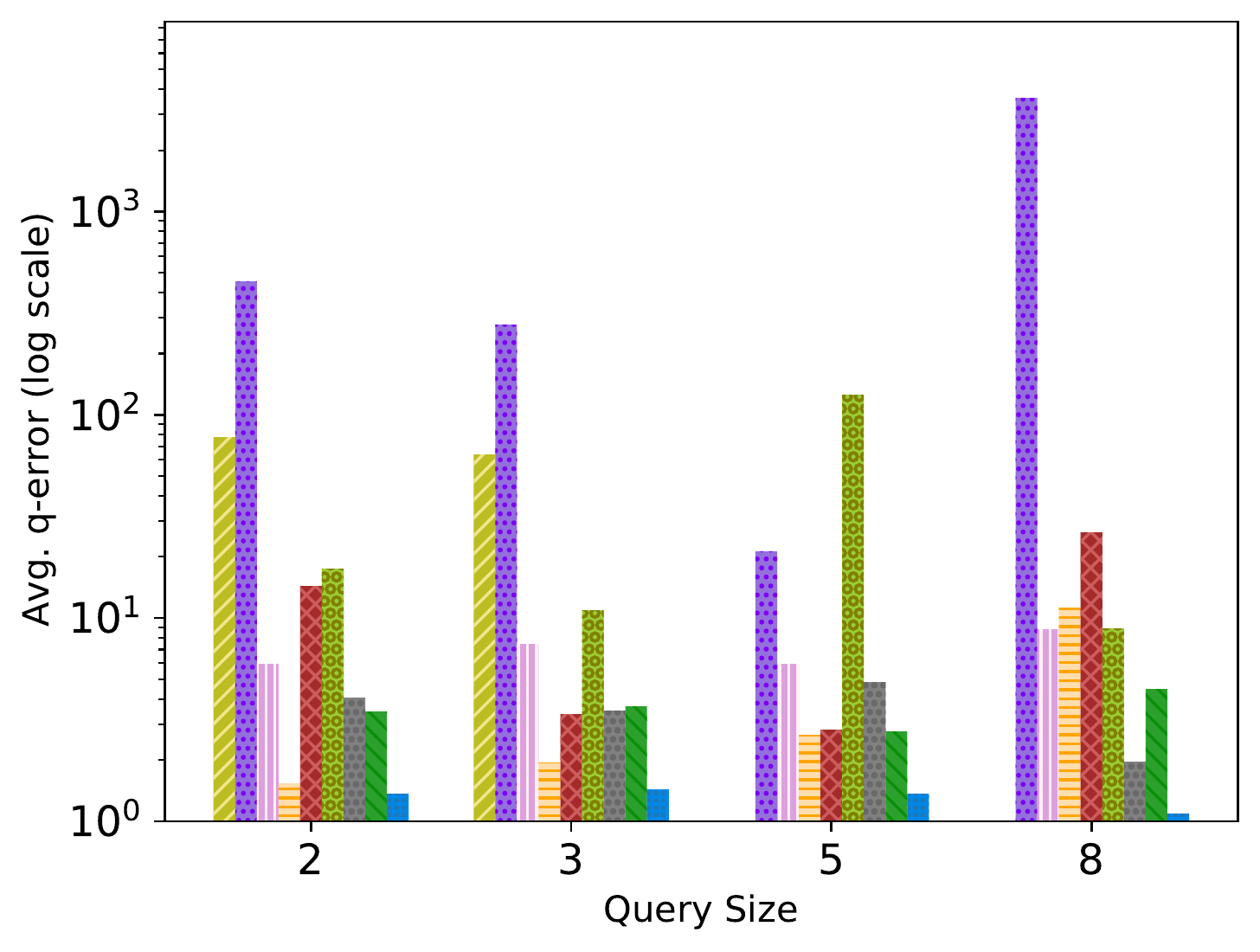}
    \label{fig:avg_q-err_query_size_lubm}
}
\caption{Accuracy for query size}
\label{fig:query_size_acc}
\end{figure}

\begin{figure*}[!t]

 \center

\centering
\subfloat[SWDF]{
  \centering
  \includegraphics[width=0.31\textwidth]{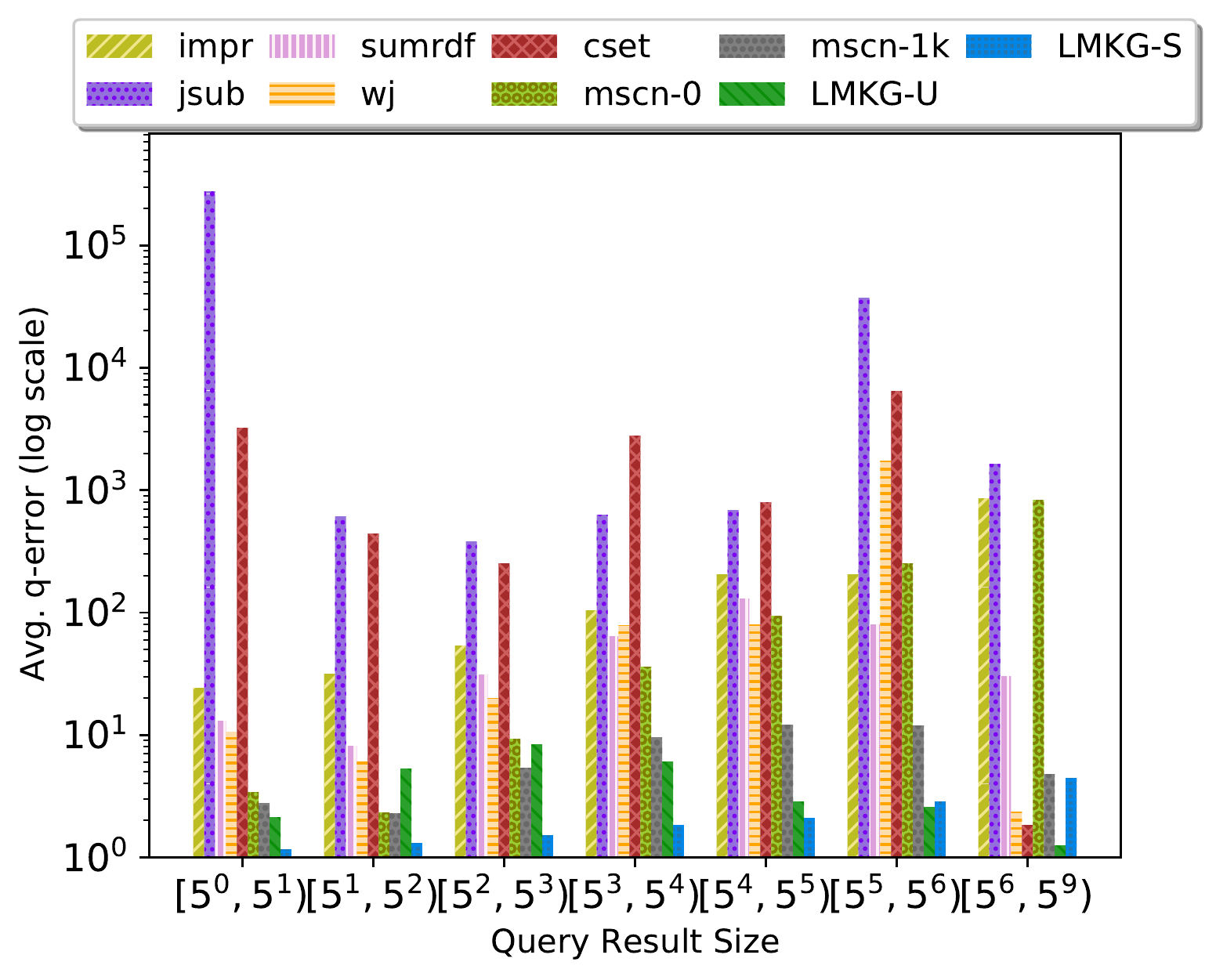}
  \label{fig:avg_q-err_range_swdf}
}%
\hspace{1pt}
\subfloat[LUBM]{
  \centering
  \includegraphics[width=0.31\linewidth]{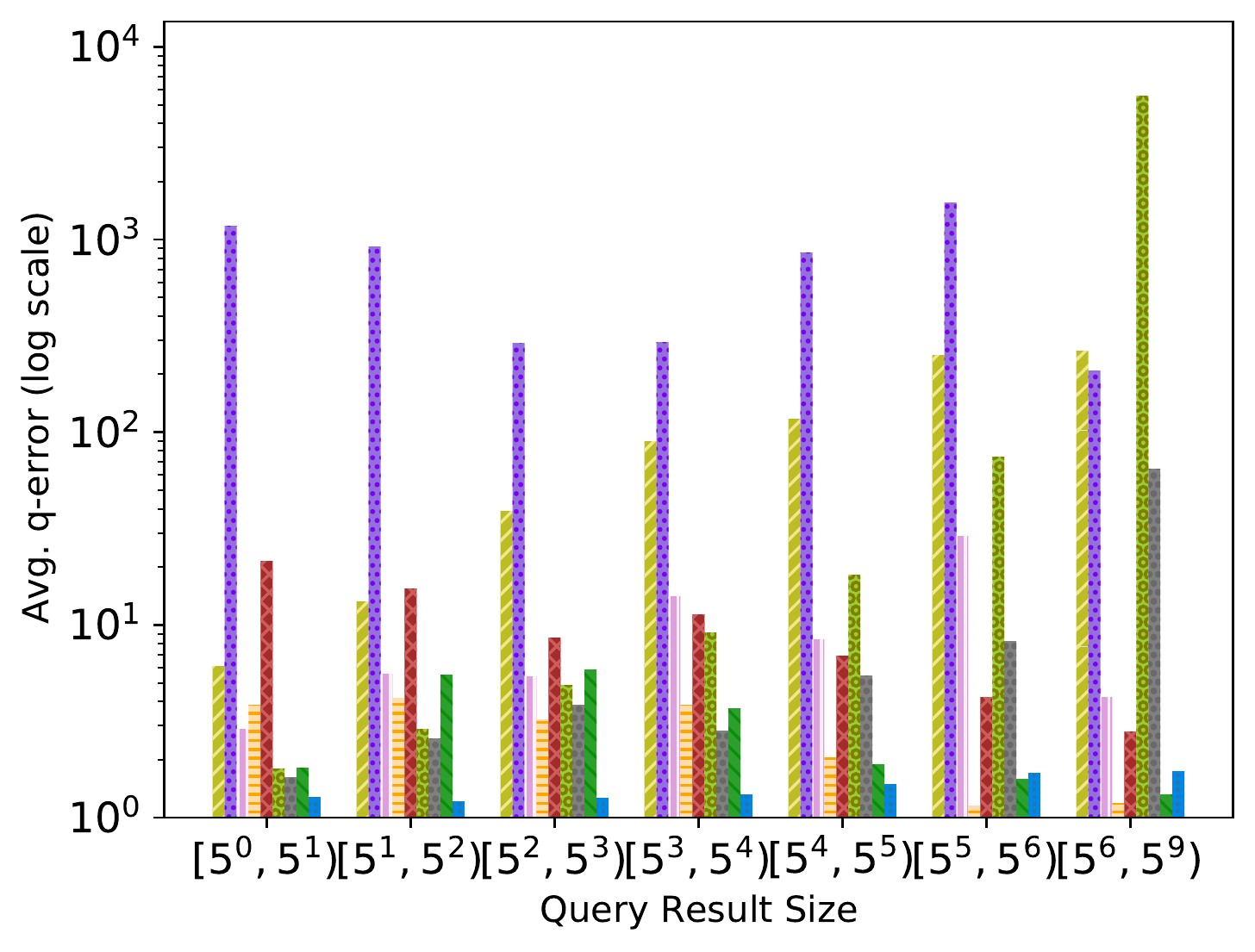}
  \label{fig:avg_q-err_range_lubm}
}%
\hspace{1pt}
\subfloat[YAGO]{
  \centering
  \includegraphics[width=0.31\linewidth]{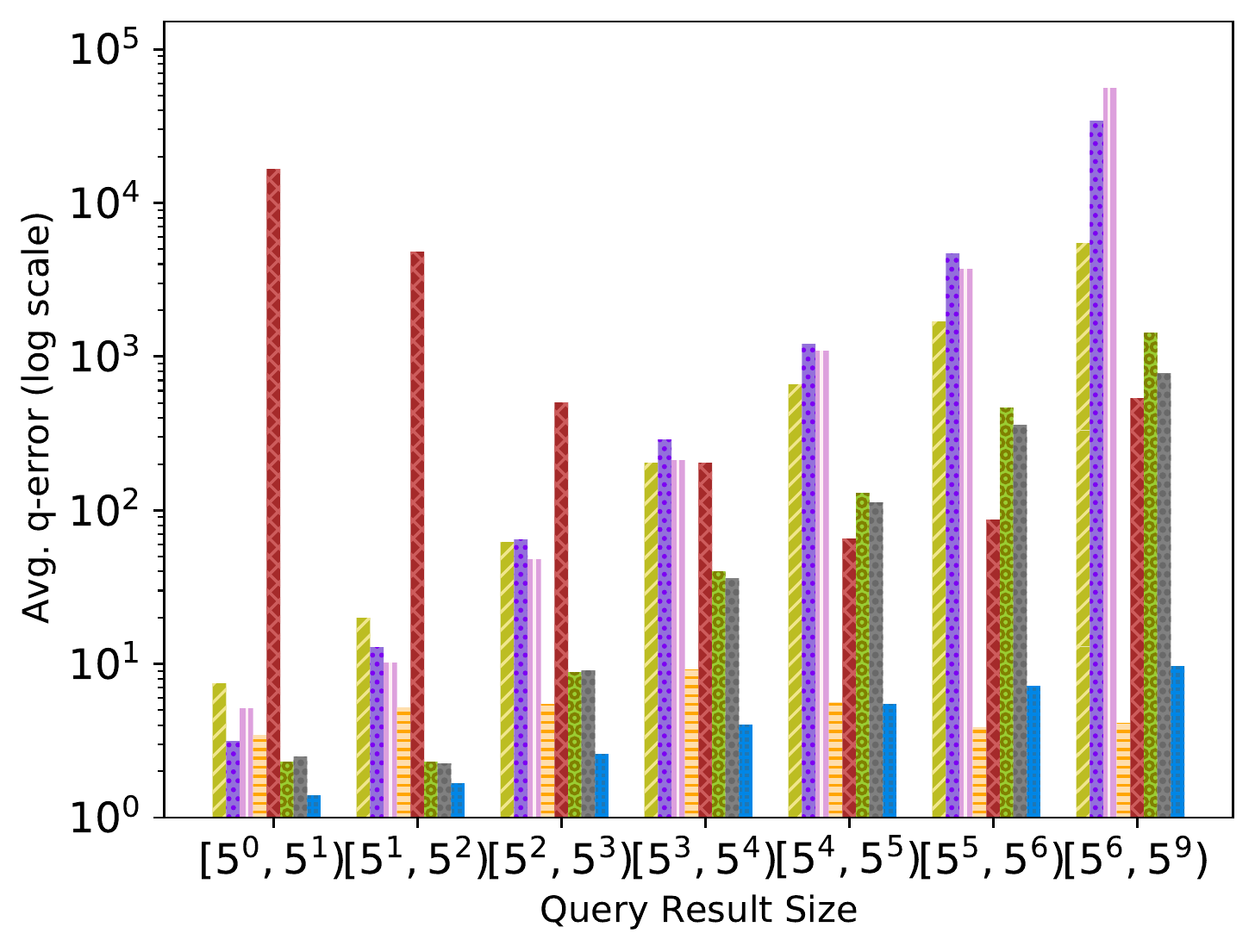}
  \label{fig:avg_q-err_range_yago}
}
\caption{Accuracy for query result size}
\label{fig:query_range_acc}

\end{figure*}

\begin{figure*}[!t]

 \center

\centering

\subfloat[SWDF]{
  \centering
  \includegraphics[width=0.31\linewidth]{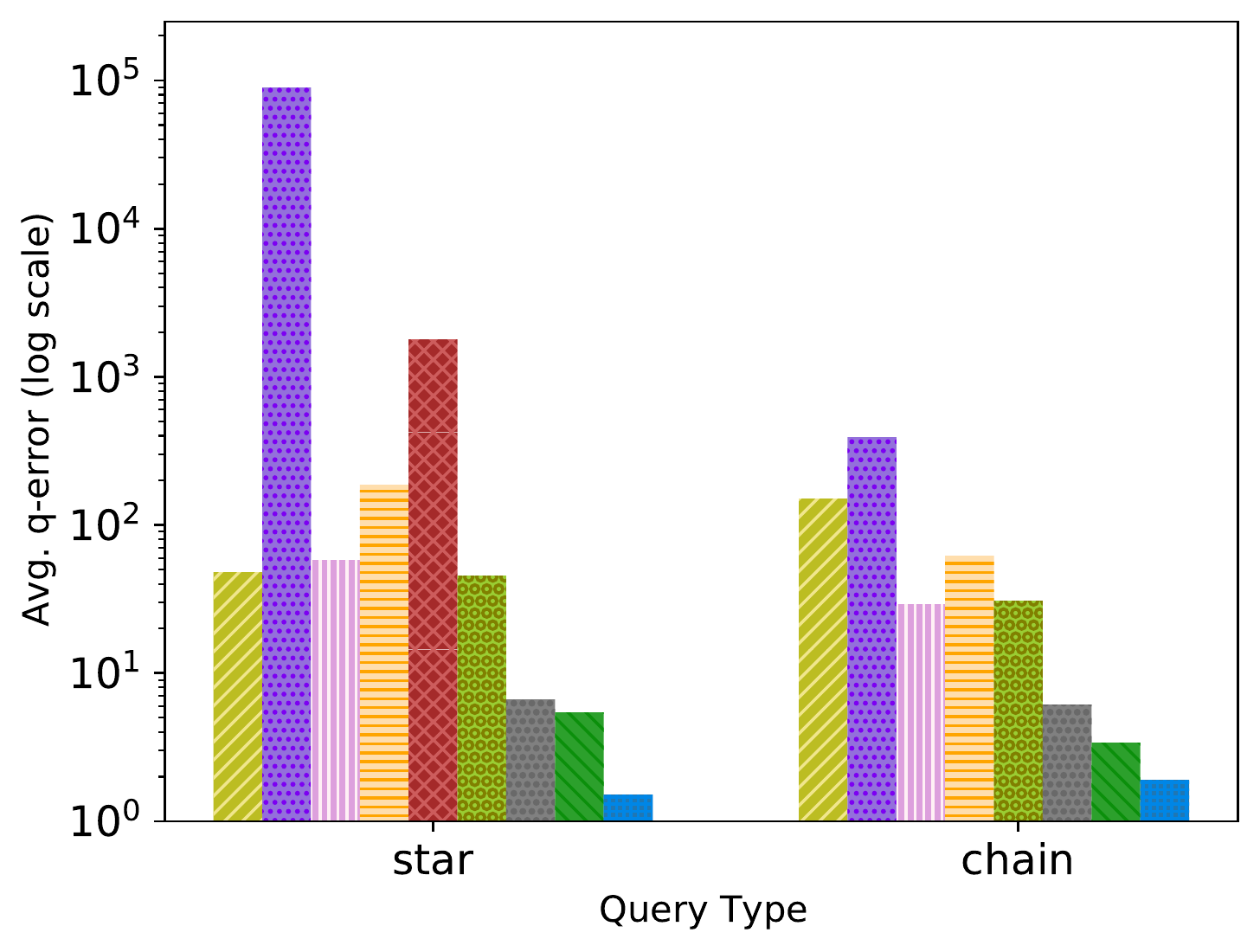}
  \label{fig:avg_q-err_query_type_swdf}
}%
\hspace{1pt}
\subfloat[LUBM]{
  \centering
  \includegraphics[width=0.31\linewidth]{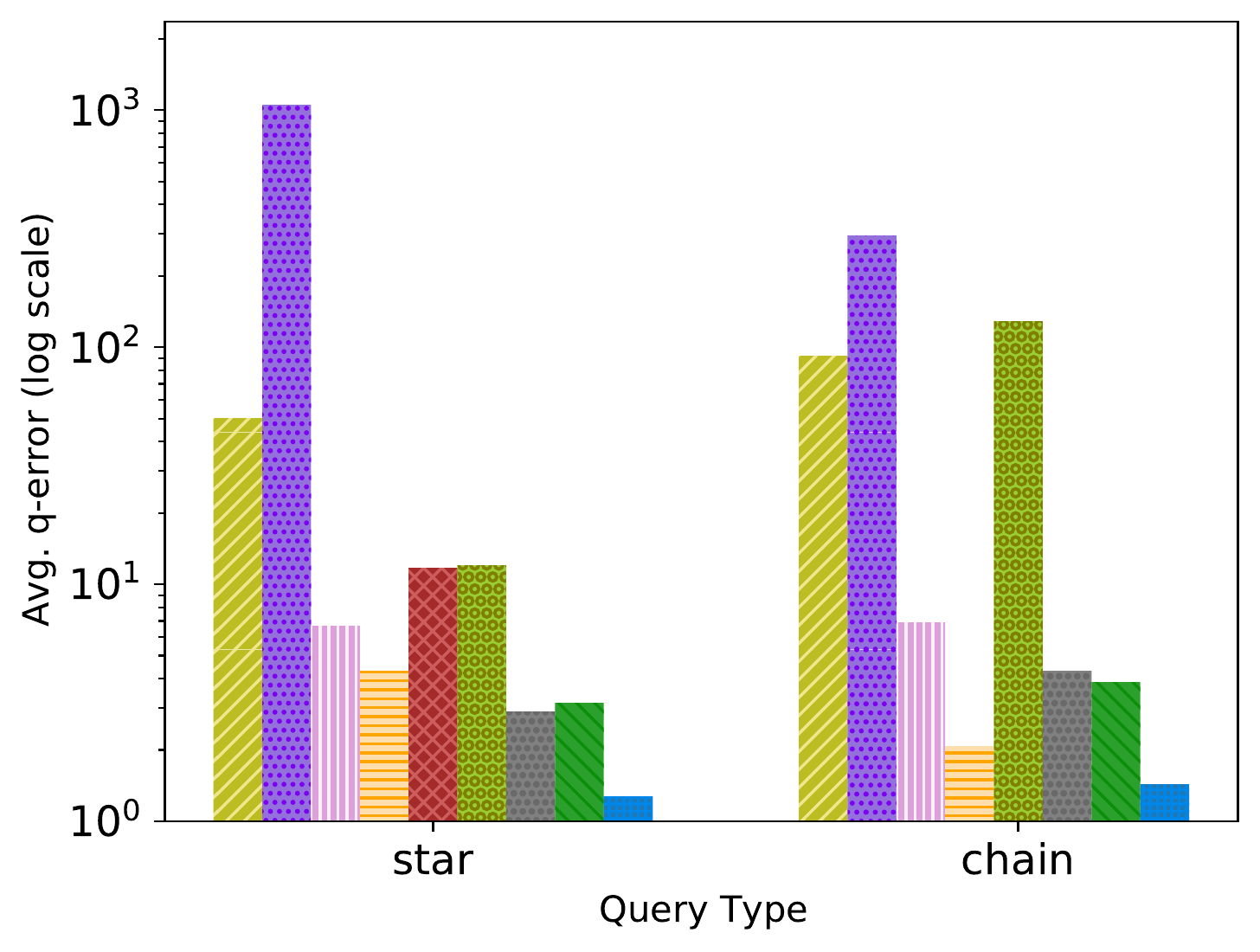}
  \label{fig:avg_q-err_query_type_lubm}
}%
\hspace{1pt}
\subfloat[YAGO]{
  \centering
  \includegraphics[width=0.31\linewidth]{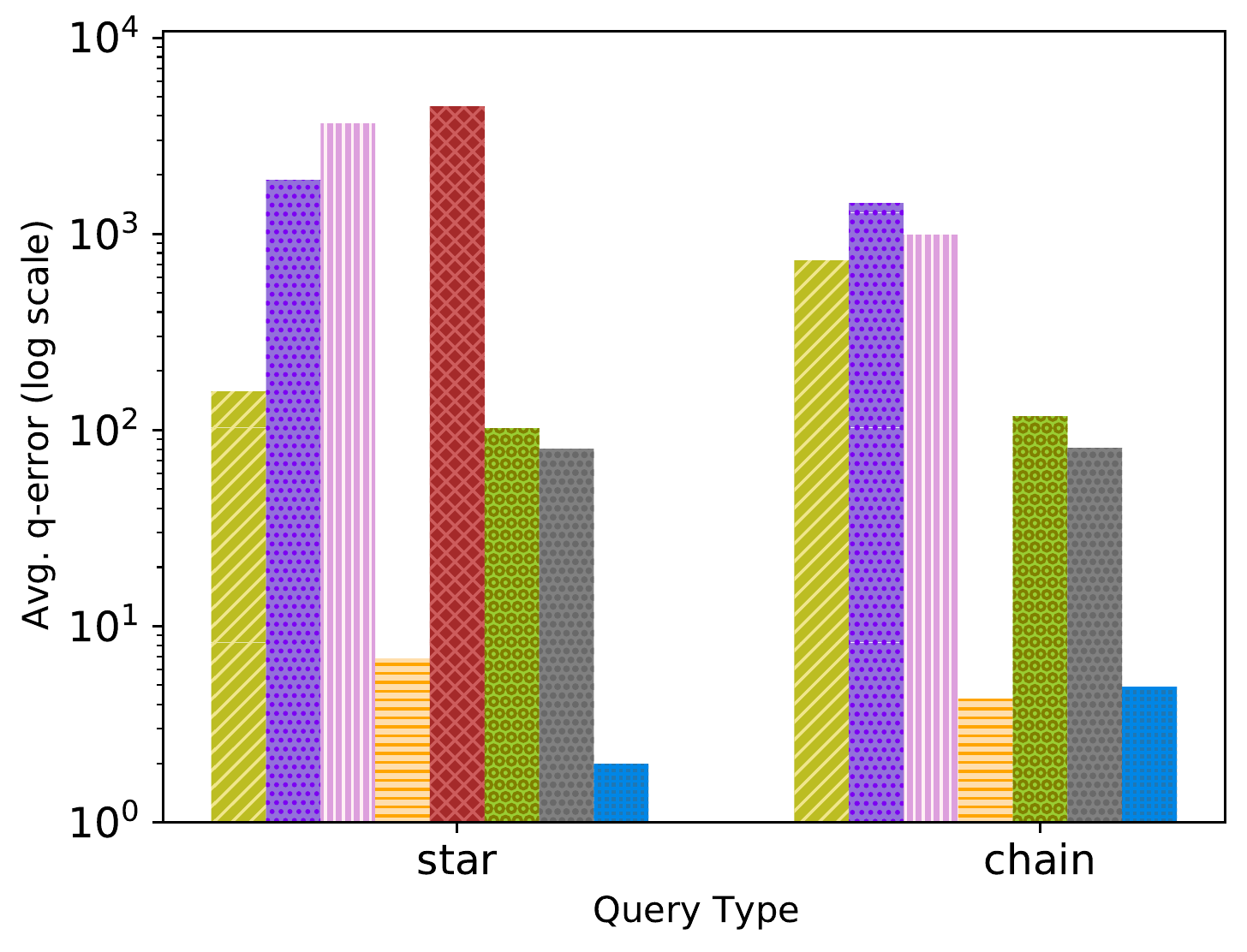}
  \label{fig:avg_q-err_query_type_yago}
}
\caption{Accuracy for query type}
\label{fig:query_type_acc}

\end{figure*}

\noindent\textcolor{brown}{\textbf{Impact of Grouping:}
In \figurename~\ref{fig:avg_q-err} we show the accuracy of the following LMKG-S models: specialized model for queries of specific type and size, model grouped by size, model grouped by type, and model for every query type and size (SingleModel). We stop after 50 epochs, where every model consists of two layers and the same configuration. 
For almost every case, the specialized model overfits the queries and produces the best estimates. The single model, as expected, has the lowest estimation accuracy. Knowing that the model trains on a much larger dataset, this accuracy can be acceptable, especially in environments with a small memory budget. The models grouped by size and type produce good estimates although worse than the specialized model. Having this accuracy comparison in mind and since the number of specialized models needed for the various combinations is significantly higher compared to the grouped models, for the experimental analysis we chose the query size grouping.}

\noindent\textbf{Impact of Outliers:} Upon measuring the models' accuracy, it is evident that unlike LMKG-U whose accuracy is impacted by the data dimensionality and range domains of the terms, LMKG-S is extremely affected by outliers. Therefore, in \figurename~\ref{fig:outlierImpact} we measure the influence of outliers in star queries, where the impact of outlier removal is most evident.
We can see that even if we remove the top-10 outliers from the query data, we achieve a higher accuracy of the model. This trend continues when a larger fraction of the outliers is removed. Although we apply normalization and scaling of the data, the impact of the skewness is still evident. Therefore, given a larger space budget, a possible improvement can be to store the cardinalities of the outliers on the side. For a fair comparison with the competitors, we proceed without this improvement.

\vspace*{-1.5mm}
\subsection{Comparison with Competitors}
\label{ssec:comparison_with_competitors}
\textcolor{brown}{For the comparison with the competitors, for LMKG-S, we used SG-Encoding and query size grouping. For LMKG-U, we used pattern-bound encoding with 32-dimensional embeddings, hence, query size and type grouping.} We depict the accuracy over the different query types, query sizes, and the query result sizes for all approaches. 

\vspace*{-2mm}
\subsubsection{Accuracy Analysis}
\hfill
\newline
\textbf{Varying Query Size:}
\figurename~\ref{fig:query_size_acc} depicts the accuracy of the approaches when varying the number of joins in the queries, represented through the average q-error.
Unlike the others, whose accuracy declines for a larger number of joins,  LMKG-S is not impacted by this factor. LMKG-U has a slight degradation of the performance when the number of joins increases, due to two factors. First, for a larger number of joins, hence terms, LMKG-U needs to learn more term correlations. Secondly, the accuracy is also impacted by the quality of the sample for training, which for larger queries can be still a challenging task. 
When looking at the different query sizes, LMKG-S performs better than the other approaches, whereas LMKG-U is more accurate for SWDF than for LUBM.

\vspace*{-2mm}
\noindent\textbf{Varying Query Result Size:}
\figurename~\ref{fig:query_range_acc} shows the accuracy for different query result sizes. Each range contains the same number of queries except for the last ones, where the patterns are sparse. To depict the downsides of LMKG we include the outliers in the query distribution.
The last buckets are grouped for larger ranges involving the outliers. When varying the query result size, we can most clearly see where the LMKG-S approach fails. LMKG-S is highly prone to outliers, visible for the higher ranges. Thus, LMKG-S is mainly impacted by the skewness. LMKG-U produces more constant results throughout the ranges. However for larger datasets, due to sampling reasons, LMKG-U fails to capture the interdependencies between the rarely occurring terms. This is more evident in the smaller ranges. Hence, for YAGO, or datasets with similar properties, where the number of term values is especially large, we advise the usage of LMKG-S.  Suitable for relational data, MSCN represents the predicate values with a single feature. However, this is not adequate for large domain values, especially for larger ranges. MSCN-1K performs better, however, a sample is still not able to capture the KG's query diversity. On the contrary, our approaches give emphasis on the term values and patterns and provide better estimates. 
When comparing with the existing KG approaches, overall, LMKG-S is always better for smaller ranges, followed by LMKG-U, WJ, and MSCN-1k. CSET and WJ perform better for larger query result sizes, however, they are inferior for smaller ranges. Regarding the overall performance when considering the query result sizes LMKG-U produces the best accuracy out of all the methods.

\noindent\textbf{Varying Query Topology:}
\figurename~\ref{fig:query_type_acc} reports the accuracy specific to star and chain queries of different sizes. LMKG-S and LMKG-U almost always perform best for both query types. WJ and MSCN-1k perform well and in some cases even outperform LMKG-U. As depicted, LMKG-U is both affected by the type of the query and the datasets and their domain values for the terms. For instance, SWDF has a higher number of unique values in the terms of the star queries and thus, the overall accuracy is slightly worse than for chain queries.

\subsubsection{Memory}
\begin{table}[!t]
  \caption{Memory consumption of different approaches}
  \centering
  \renewcommand{\arraystretch}{1.6}
\scalebox{0.65}{
  \begin{tabular}{|p{1cm}|c|c|c|c|c|c|c|c|c|}
    \hline

{\textbf{Dataset}} & \multicolumn{3}{c|}{\textbf{LMKG-U}} & \multicolumn{3}{c|}{\textbf{LMKG-S}} & \textbf{SUMRDF} & \textbf{CSET} & \textbf{MSCN}\\
    \cline{2-10}
    & \textbf{$k=2$} & \textbf{$k=3$} & \textbf{$k=5$} & \textbf{$k=2$} & \textbf{$k=3$} & \textbf{$k=5$}  & \textbf{Complete}  & \textbf{Complete} & \textbf{0/1K} \\
    \hline
    SWDF & 19MB & 43MB & 46MB & 4MB &  4MB  & 8MB & 1.2MB  & 816.7 KB & 5 / 8 MB \\ \hline
    LUBM & 80 MB & 78MB & 27MB & 2MB & 2MB  & 5MB  & 8.8MB  & 8.6 KB & 5 / 8 MB \\ \hline
    YAGO & X & X & X & 4MB  & 7MB  & 7MB & 342MB  & 5MB & 5 / 8 MB  \\ \hline
  \end{tabular}
}
\label{table:memory-consumption}
\vspace*{-4mm}
\end{table}

For the memory consumption, we compare with the summary-based approaches and MSCN. 
We measure the complete model for LMKG-U and LMKG-S. 
Although for LMKG-U and LMKG-S a model for a specific size $k$ can answer smaller queries we make the following table for completeness.
Intuitively, sample-based approaches have an advantage since they use the KG for drawing samples.
In Table~\ref{table:memory-consumption} we show the results for the different KGs. 
LMKG-S provides smaller memory than LMKG-U as well as some of the competitors. MSCN-0 has a smaller footprint due to the much smaller input, with the cost of worse accuracy.
CSET is better for SWDF and LUBM, however, for YAGO it has a larger size. On the contrary, the memory consumption of LMKG-U increases with the dataset size and the number of unique terms involved. In some scenarios, LMKG-U requires less memory because for large query sizes the scarcity of patterns results in fewer unique terms contributing to a smaller sample size and a model with fewer layers and neurons. 

\subsubsection{Estimation Time}
\begin{figure}[!t]
\centering
\subfloat[Varying query size (SWDF)]{
  \centering
  \includegraphics[width=0.47\linewidth]{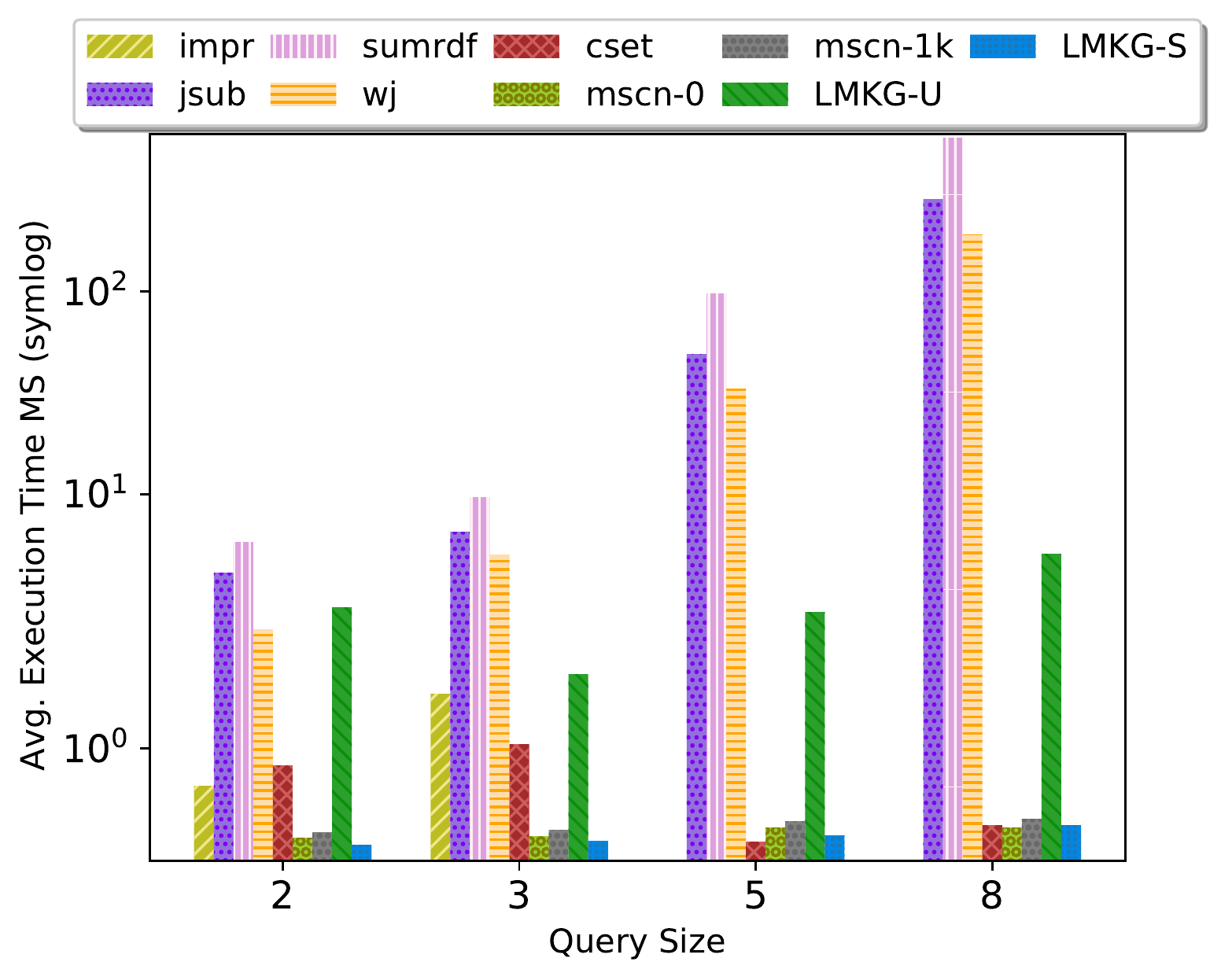}
  \label{fig:exec_time_q_size_swdf}
}%
\hfill
\subfloat[Varying query type (SWDF)]{
  \centering
  \includegraphics[width=0.47\linewidth]{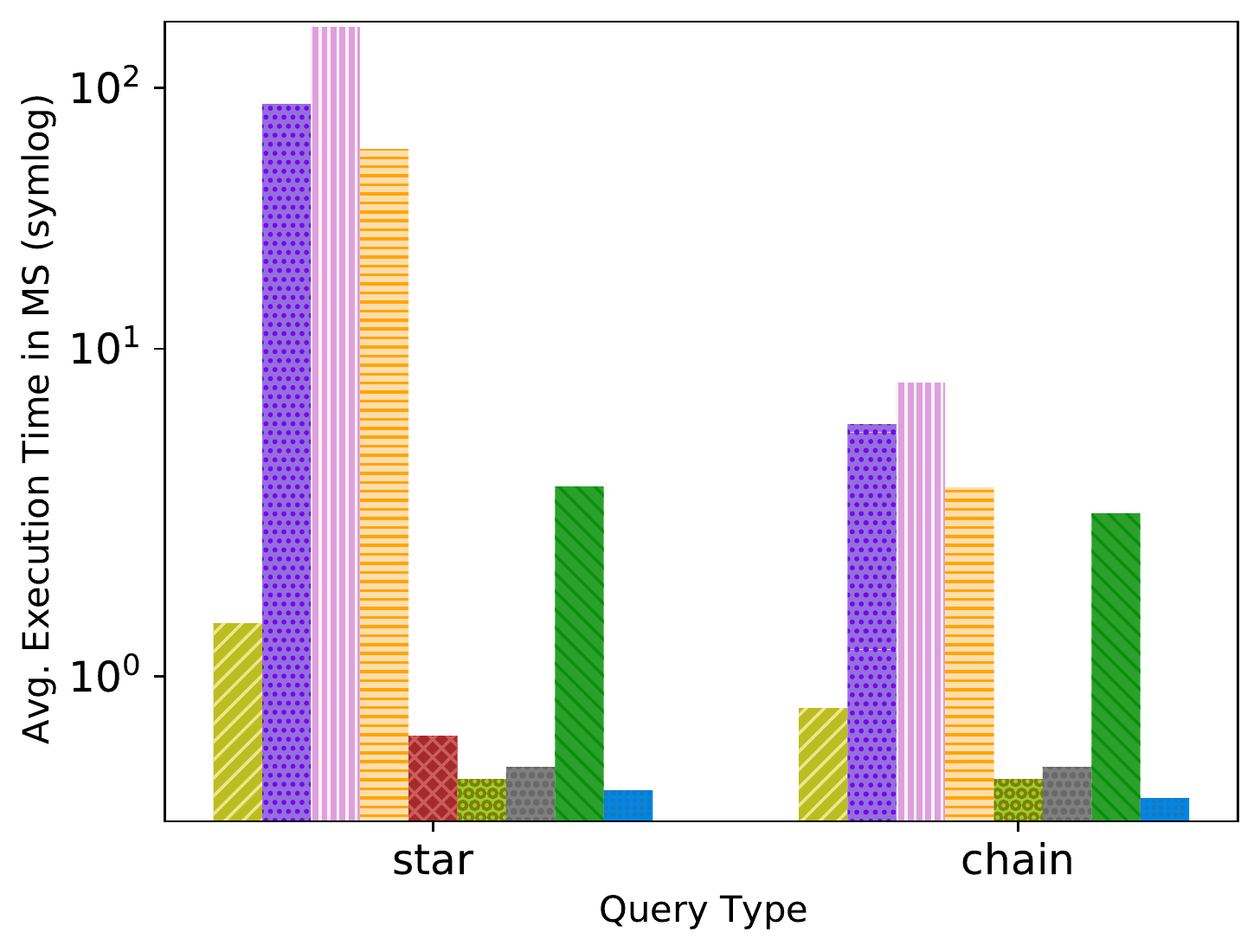}
  \label{fig:exec_time_q_type_swdf}
}%
\hfill
\subfloat[Varying query size (LUBM)]{
  \centering
  \includegraphics[width=0.47\linewidth]{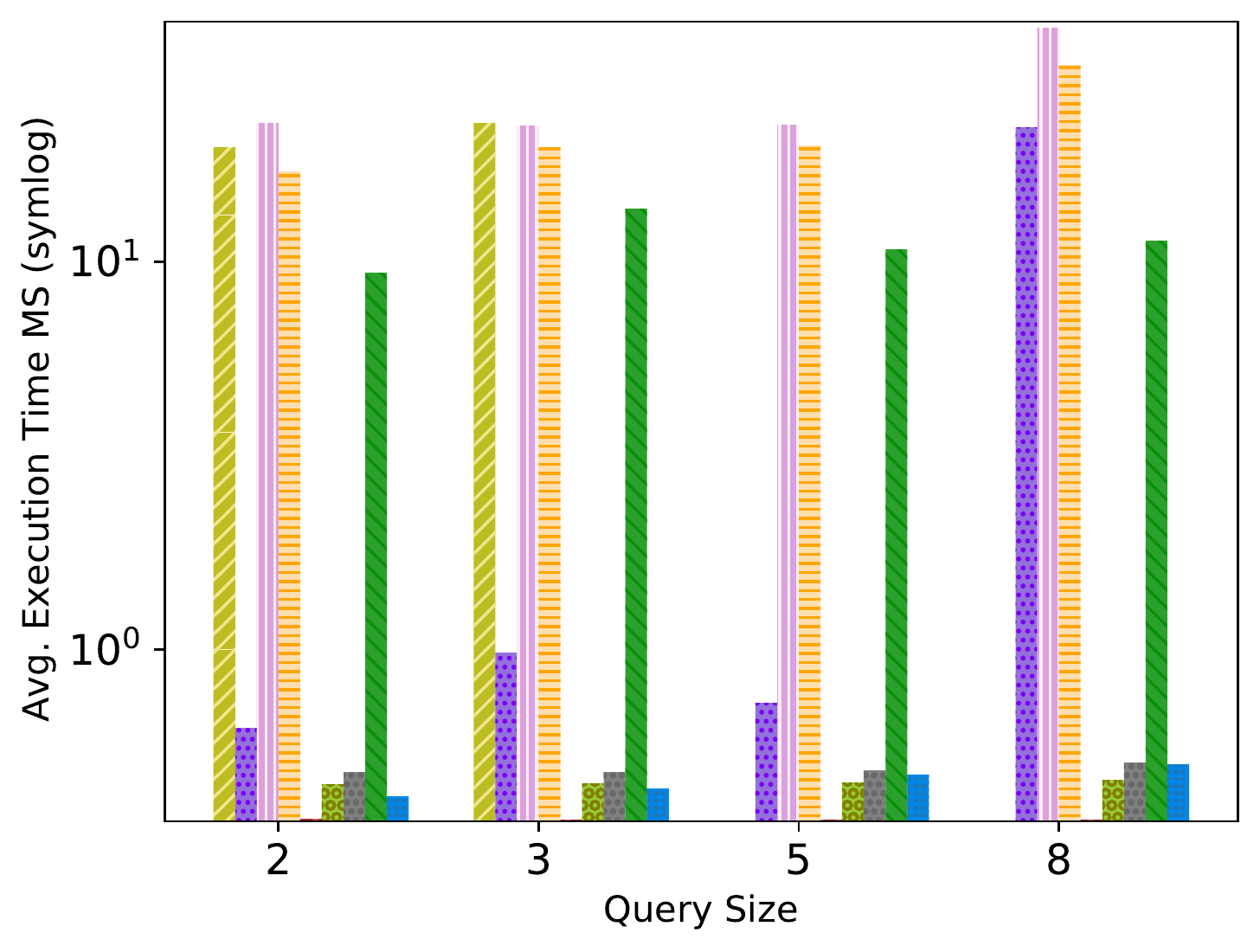}
  \label{fig:exec_time_q_size_lubm}
}%
\hfill
\subfloat[Varying query type (LUBM)]{
  \centering
  \includegraphics[width=0.47\linewidth]{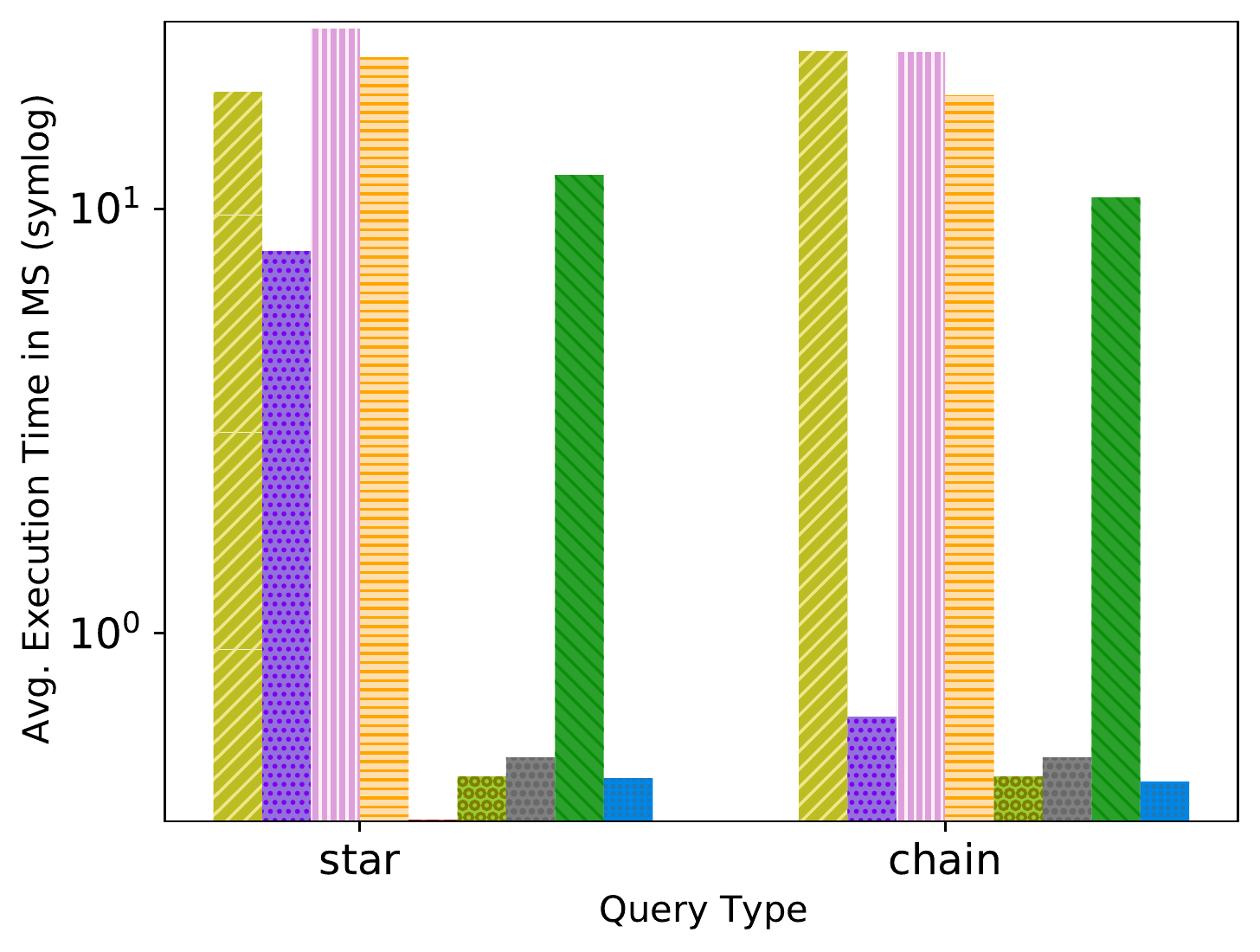}
  \label{fig:exec_time_q_type_lubm}
}
\caption{Estimation time}
\label{fig:estimation_time}
\vspace*{-4mm}
\end{figure}

In \figurename~\ref{fig:estimation_time} we show the estimation time for the different types of queries varied by their size. For the sampling approaches, we measure the time of generating 30 samples since G-CARE needs 30 samples for producing an accurate final estimate. A smaller sample size produced much worse accuracy, but intuitively, a faster approach. Intuitively MSCN has a similar prediction time as LMKG-S. As shown, LMKG-S performs better than all the approaches, except for the CSET approach. LMKG-U is as good or sometimes better than the other approaches. Other than CSET, the other approaches are majorly affected by the query size.

\vspace*{-2mm}
\subsection{Lessons Learned}
\label{ssec:lessons_learned_conclusion}
\textbf{Challenges and Future Directions:} The main performance degradation in LMKG-S is not a result of the complexity of the queries, but of the large outliers. As shown by the experiments, LMKG-S overall outperforms every approach, however, it fails to accurately estimate on outliers. To overcome this problem, we suggest the usage of a buffer list for storing the outliers. Differently, LMKG-U provides a more constant performance. However, the main difficulty for this model comes from a large number of unique term values, causing a high number of correlations for which the current RW sampling has not proved efficient in all of the cases. Future work involves exploring a more optimal sampling approach and a reduction of the per term dimensionality for LMKG-U.
Parallel to LMKG, the use of a deep learning model, such as GraphAF~\cite{DBLP:conf/iclr/ShiXZZZT20}, may be beneficial for sampling from the knowledge graph.

\textbf{Optimal Use-cases:}
Evidently, a generalization for different sizes or types of queries requires a higher training time than creating the other approaches. Considering all the aspects of query cardinality estimation in knowledge graphs, LMKG is optimal for scenarios where a workload is given or a cardinality for a specified range of queries (e.g., up to $k$ joins) is wanted.
Additionally, the learned models in the LMKG framework are practically useful when considering query optimization, where a reordering of different patterns of smaller sizes is needed. Although the models are more accurate and can be useful for query cardinality estimation of larger or combined queries, their scaling depends on the training data, and thus, a sampling-based approach or a combination of sampling and learned approach may be more efficient.

\pdfoutput=1

\section{Conclusion}
\label{sec:conclusion}
We addressed the problem of applying deep learning methods for cardinality estimation in KGs by utilizing both, supervised and unsupervised deep learning models. 
To efficiently feed knowledge subgraphs to our models, we investigated various encodings and introduced a novel encoding that is especially useful for our supervised model. By focusing on the subgraph patterns that constitute the KG, our encoding drastically reduces the input size and enables us to train the  models on more than one query type. We additionally explained possible improvements to our framework and future directions. Through the experimental evaluation, we showed that LMKG-S and LMKG-U exceed the state-of-the-art approaches in terms of accuracy while keeping a small memory footprint and requiring less time for generating the estimates.

\end{document}